\documentstyle[epsfig,rotate,12pt]{article}
\newlength{\dinwidth}  
\newlength{\dinmargin}
\def\bsll{$b \rightarrow s \ell^+ \ell^- $ }
\def\bslll{$b \rightarrow s \ell^+ \ell^- $}
\def\bxsll{$B \rightarrow X_s \ell^+ \ell^- $ }
\def\bxslll{$B \rightarrow X_s \ell^+ \ell^- $}

\def\bxsg{$B \rightarrow X_s \gamma $ }
\def\bsg{$b \rightarrow s \gamma $}
\def\absvcb{\left| V_{cb} \right|}

\def\s{\hat{s}}
\def\u{\hat{u}}
\def\z{v \cdot \hat{q}}
\def\cnt{C_9^{\mbox{eff}} \mp C_{10}}
\def\ml{\hat{m}_l}
\def\ms{\hat{m}_s}

\def\mc{\hat{m}_c}
\def\mb{\frac{M_B}{{m_b}^3}}
\def\lo{\hat{\lambda}_1}
\def\lt{\hat{\lambda}_2}
\def\loo{\lambda_1}
\def\lto{\lambda_2}
\def\q{\hat{q}}
\title{Power Corrections in the
 Decay Rate and Distributions in \bxsll in the Standard Model }
\author{\vspace{1cm}\\
	{\bf A. Ali}\thanks{E-mail address: ali@x4u2.desy.de} , 
	{\bf G. Hiller}\thanks{E-mail address: ghiller@x4u2.desy.de} , \\
	 Deutsches Elektronen-Synchroton DESY, Hamburg \\
	\vspace{5mm}\\
	{\bf L.T. Handoko}\thanks{On leave from P3FT-LIPI,
	Indonesia. E-mail address: 
        handoko@theo.phys.sci.hiroshima-u.ac.jp} ,
        and {\bf T. Morozumi}\thanks{E-mail address:
        morozumi@theo.phys.sci.hiroshima-u.ac.jp} \\
	Department of Physics, Hiroshima University \\
	1-3-1 Kagamiyama, Higashi Hiroshima  -  739, Japan\\}
\date{}

\begin{document}
\setlength{\baselineskip}{24pt}

\maketitle
\begin{picture}(0,0)
       \put(325,350){DESY 96-206}
       \put(325,335){HUPD-9615}
       \put(325,320){September 1996}
\end{picture}
\vspace{-24pt}

\setlength{\baselineskip}{8mm}

\begin{abstract}
We investigate the leading power corrections to the decay rates
and distributions in the decay \bxsll
in the standard model (SM) using heavy quark expansion (HQE) in $(1/m_b)$
and a phenomenological model implementing the Fermi motion effects of the
$b$ quark bound in the $B$ hadron. In the HQE method, 
we find that including the leading power corrections the decay width
$\Gamma(\mbox{\bxslll})$ decreases by about $4\%$ and the
branching ratio ${\cal B}$(\bxslll) by about $1.5\%$ from their (respective)
parton model values. The dilepton invariant mass spectrum is found to 
be stable against power corrections over a good part of this spectrum.
However, near the high-mass end-point this distribution becomes negative 
with the current value of the non-perturbative parameter $\lambda_2$ (the 
$\lambda_1$-dependent corrections are found to be inoccouos), implying
the breakdown of the HQE method in this region. Our results are at variance
with the existing ones in the literature in both the decay rate and the 
invariant dilepton mass distribution calculated in the HQE approach.
As an alternative, we implement the non-perturbative effects in the decay
\bxsll using a phenomenologically motivated Gaussian Fermi motion model.
We find small corrections to the branching ratio, but the non-perturbative 
effects are perceptible in both the dilepton mass distribution and the
Forward-Backward asymmetry in the high dilepton mass region. Using this 
model for estimating the non-perturbative effects, modeling the dominant
long distance (LD) contributions from the decays
$B \to X_s +(J/\psi, \psi^\prime,...) \to X_s \ell^+ \ell^- $, and taking
into account the next-to-leading order perturbative QCD corrections in
$b \to s \ell^+ \ell^-$, we present the decay rates and 
distributions for the inclusive process \bxsll in the SM.
\end{abstract}

\thispagestyle{empty}
\newpage
\setcounter{page}{1}

\section{\bf Introduction}
Rare $B$ decays \bxsll and \bxsg are well suited to
test the SM and search for physics beyond the SM.
 In the SM, such processes are
governed by the Glashow-Iliopoulos-Maiani (GIM) mechanism \cite{GIM},
and their rates and distributions  are sensitive to the
top quark mass and the Cabibbo-Kobayashi-Maskawa (CKM) matrix elements
\cite{CKM}.
 First measurements of
the decay rates for the exclusive decay $B \to K^* + \gamma$ 
\cite{CLEOrare1} and the inclusive decay
\bxsg \cite{CLEOrare2} have been reported by the CLEO collaboration.
 At the partonic
level, the complete leading order (LO) anomalous dimension matrix involving
the  \bsg ~decay
was calculated in \cite{Ciuchini}-\cite{Ciuchini94} in the context of an 
effective five-quark theory. First calculations of
the gluon bremsstrahlung and virtual corrections, which are part of the 
next-to-leading-order perturbative QCD  
improvements, were reported in \cite{ag1}-\cite{Pott} (see also 
\cite{Sterman94,Dikeman}).
The NLO virtual corrections to the matrix elements have been completed 
in \cite{GHW96}. A first calculation of the hitherto missing
NLO anomalous dimension matrix has been recently reported 
in \cite{Misiak96}. Leading
power corrections in $(1/m_b)$ to the decay rate $\Gamma(\mbox{\bxsg})$
(and $\Gamma(B \to X \ell \nu_\ell)$, which is often used to estimate the
branching ratio ${\cal B}(\mbox{\bxsg})$) have
also been calculated in the heavy quark expansion (HQE) approach
\cite{georgi,manoharwise}.
A quantitative measure of the rapport between experiment
and present estimates in the SM is the CKM matrix element ratio
$\vert V_{ts}\vert /\vert V_{cb} \vert$ for which a value 
$\vert V_{ts}\vert /\vert V_{cb} \vert =0.85 \pm 0.12 ~\mbox{(expt)} \pm 
0.10 ~\mbox{(th)}$ has been obtained from the inclusive decay rate for \bxsg 
\cite{ali96}, in agreement with the bounds obtained from unitarity
of the CKM matrix \cite{PDG96}.

\par
 It is known that the inclusive energy spectra in the 
decays $B \to X \ell \nu_\ell$ and \bxsg are not entirely
calculable in the HQE framework 
\cite{manoharwise},\cite{Bigietal2}-\cite{JR94}. In particular, the
end-point energy spectra are problematic in that the energy released for
the light quark system in the decay $b \to q X$ (here $X=\gamma$ or a 
dilepton pair) is not of order $m_b$ but
of order $\bar{\Lambda}$, where
$\bar{\Lambda} = m_B - m_b =O(\Lambda_{\mbox{\small QCD}})$. Hence, 
the expansion parameter in the HQE approach, which is formally
of $O(1/Q^2) =O(1/m_b^2)$, near the end-point  gets replaced by a quantity 
which is of $O(1/K^2)=O(1/\bar{\Lambda}m_b) ~\gg ~O(1/m_b^2)$,
implying the onset of the breakdown of the HQE method. To make contact with
experiments one has to smear the energy spectrum in question
over an energy interval sufficiently larger than $\Lambda_{\mbox{QCD}}$.
Thus, direct comparison of theoretical distributions with experiments
requires additional input in terms of phenomenological   
models, e.g., the Gaussian Fermi-motion model of \cite{aliqcd},
which incorporate such smearing.
The smearing effects are very important in \bxsg \cite{ag1,ag2,Dikeman}  
 and not negligible in
the lepton and hadron-energy spectra in the decays $B \to X_c (X_u) \ell
\nu_\ell$, either \cite{aliqcd}-\cite{Greub96}.
Alternatively, one may have to resort to a 
resummation of the power corrections near the
end-point \cite{Bigietal2,neubertbsg}. Such resummations, however, remain 
so far inconclusive.

 \par
In this paper, we address the related FCNC process \bxslll, $\ell 
=e,\mu$. (Since we neglect the lepton masses in our calculation, our
results are not applicable to the decay $B \to X_s \tau^+ \tau^-$.)
The SM-based short distance (SD) contribution to the decay rate for the 
partonic process \bslll, calculated in the free quark decay
approximation, has been known in the LO approximation for some time 
\cite{Grinstein89}. In the meanwhile, also the NLO perturbative QCD 
corrections  have been
calculated which reduce the scheme-dependence of the LO effects in these 
decays \cite{Misiak1,buras}. In addition, long distance (LD) effects, which 
are expected to be very important in the
decay \bxsll  \cite{long}, have been estimated from data on the premise 
that they arise dominantly from  
the charmonium resonances $ J/\psi$ and $\psi'$ through the decay chains
$B \rightarrow X_s J/\psi (\psi') \rightarrow X_s \ell^+ \ell^-$. Higher
resonances ($\psi'', \psi''',...)$ also contribute though at a reduced
level.
Estimates of the LD effects away from the resonance 
regions involve specific assumptions about the $q^2$-dependence of the
relevant vertices, which at present can only be obtained in specific 
models \cite{long} - \cite{Ahmady96}.

\par
The particular aspect we are interested in is an
estimate of the non-perturbative effects on the decay distributions
in \bxsll , which take into account the $B$-hadron wave function effects
and incorporate the physical threshold in the final state on the underlying
partonic calculations. This effects both the SD- and LD-contributions,
and to the best of our knowledge has not yet been
calculated. Closely related to this aspect is the question of 
power corrections to the parton model decay rates and spectra which 
have been calculated for the  SD-part of the dilepton 
invariant mass distribution in 
\bxsll by Falk, Luke and Savage \cite{falketal} (henceforth this 
paper is referred to as FLS) using the HQE approach.
We reevaluate these corrections in this paper, reaching
different results and conclusions than in the FLS paper which we specify 
later.

 From the power corrections calculated in the HQE approach
 for the decays  \bxsg  and  $B \to X \ell \nu_\ell$ 
 \cite{georgi,manoharwise}, we recall that there are no leading, i.e.,
$O(1/m_b)$, corrections in the inclusive rates.
Likewise, in the decay \bxslll, the first non-vanishing corrections
to the inclusive rates are of $O(1/m_b^2)$. Furthermore,
the dilepton mass spectrum in \bxsll is found to 
be well behaved in the HQE framework in the
{\it entire} dilepton mass range in FLS \cite{falketal}. In particular, the 
high dilepton invariant mass spectrum in the parton model
is found to receive moderate power 
corrections, typically $O(10\%)$, increasing the dilepton yield in
\bxsll (see Fig.~2 in \cite{falketal}).
This result differs qualitatively from analogous power corrections in
the lepton energy spectra in $B \to X \ell \nu_\ell$, which are large and 
negative near the end-points (see, for example, Figs.~5 - 8 in the
paper by Manohar and Wise \cite{manoharwise}). In addition,
taking the $V-A$ limit in the matrix element for \bxslll, the 
differential distributions and decay rate in this process can be related to
the corresponding quantities in the semileptonic decay $B \to X \ell 
\nu_\ell$. The power corrections in the latter decays have been calculated
and discussed at great length by Bigi et al.~in \cite{georgi} and by
Manohar and Wise \cite{manoharwise}. We are of the opinion that both
the power corrected dilepton spectrum and the inclusive decay rate
$\Gamma(\mbox{\bxslll})$ obtained by integrating
this spectrum in FLS are at variance with the results in 
\cite{georgi,manoharwise} in this limit (see Appendix C).
 In view of the impending interest in the decay \bxslll,
in particular the dilepton mass spectrum and the Forward-backward asymmetry 
involving $\ell^+$ versus $\ell^-$ \cite{amm91},
 which have been put forward as precision test of the SM in
the FCNC sector and hence a possible place for discovering new physics
\cite{agm94,choetal},
we have recalculated the power corrections in this process in the SM using 
the HQE approach.

To that end, we have computed the Dalitz distribution,
${\rm d}^{2}{\cal B}/{\rm d}\hat{s} {\rm d}\hat{u}$, for the decay \bxsll 
(see section 2 for the definition of these variables), taking
into account the NLO perturbative QCD correction in $\alpha_s$ and the
leading $1/m_b$ corrections in the HQE approach. In doing this, we have 
also kept the $s$-quark mass effects.
Integrating over one of the variables, the
resulting expressions for the dilepton invariant mass and the
FB asymmetry are derived.
While the power-corrected FB asymmetry in \bxsll is a new result, not
presented earlier, our expression for the 
power-corrected dilepton mass distribution is not in agreement with the one 
presented in FLS \cite{falketal}.
 Since the derivations of the final results
for ${\rm d}\Gamma(\mbox{\bxsll})/{\rm d}\hat{s}$
and the FB asymmetry ${\cal A}(\hat{s})$ are
rather involved, we have decided to give the details of the
calculations so that they can be checked stepwise and the source of this
discrepancy pinned down accordingly. Some checks of our results in the 
limiting case mentioned above are
already possible and have been carried out. In particular, we are able
to derive the results in \cite{georgi,manoharwise} taking the appropriate 
limit of our expressions in \bxslll (see again Appendix c). 

\par
We find that the final-state distributions in \bxsll are not calculable 
entirely
in the HQE approach, as the dilepton mass distribution becomes negative in
the end-point region. While this defect may be resuscitated by resummation
of the HQE-power corrections, we do not attempt this here. Instead, we
estimate the
non-perturbative effects on the decay rates and distributions in \bxsll
by invoking the Gaussian Fermi motion model \cite{aliqcd}. This model has 
been used successfully in the analysis of the lepton energy spectrum in the
semileptonic decays $B \to X \ell \nu_\ell$ \cite{CLEOslfm} and
the  photon energy spectrum in \bxsg \cite{ag2}.
As pointed out in \cite{manoharwise} on the example of $B \to X\ell 
\nu_\ell$, this model reproduces the effect due to the kinetic energy
term $\lambda_1$ in the HQE approach, if the $b$-quark 
mass is appropriately defined, but there is no analogue of $\lambda_2$
(the matrix element of the magnetic moment operator)
in the Fermi motion model. The distributions in the two approaches (HQE 
and the Fermi motion model) are hence, in general, different, which is most
noticeable near the end-points. By construction, there are no negative 
probabilities encountered in the Fermi motion model and the final state 
thresholds can be correctly incorporated.

  This paper is organized as follows: In section 2, we
derive the double differential distribution
${\rm d}^{2}{\cal B}/{\rm d}\hat{s} {\rm d}\hat{u}$ for the decay \bxslll,
including the explicit $O(\alpha_s)$  and the
leading power corrections in $1/m_b$, giving in Appendix A
the individual contributions to the structure functions
from several contributing sources governing these decays. 
Some of the lengthy expressions obtained in the derivation of the  
HQE-improved Dalitz distribution are displayed in Appendix B. 
The power-corrected dilepton invariant mass
distribution and the FB asymmetry in \bxslll,
together with their simplified versions in the limit $m_s =0$, are also 
given in this section. We also present here numerical 
comparisons in the two  quantities of interest between
the parton model and the HQE-approaches, as well as differences between our 
result and the one in FLS \cite{falketal}. In Appendix C, we
present the limiting case of our results for \bxsll and compare them with the
existing ones in the literature \cite{georgi,manoharwise}. In 
Appendix D, we show (a peripheral result) that the
energy asymmetry defined in \cite{choetal} and the FB asymmetry 
introduced in \cite{amm91} are related.  
 In section 3, we implement the $B$-meson wave function effects and the
physical threshold  on the final state in \bxslll, using the NLO-corrected 
parton distributions and the Gaussian Fermi motion model 
\cite{aliqcd}. Since the calculation of the FB asymmetry in this
model involves some non-trivial kinematic transformations, we have
given the details in Appendix E. 
The LD-contributions in \bxsll are estimated in section 4,
using data on vector meson intermediate states $B \to V +X_s$, where
 $V=(J/\psi (1S),...,\psi(6S))$. The resulting dilepton mass spectrum and 
the FB asymmetry, including the wave-function and LD-effects,
are also presented here.
We conclude with a discussion of our results and possible improvements of 
the LD-effects in \bxsll in section 5.

\section{\bf Power corrections to the dilepton invariant mass distribution
            and FB asymmetry in \bxsll }
We start by defining the various kinematic variables
in the decay $b(p_b) \to s(p_s) + \ell^+(p_+) + \ell^-(p_-)$.  
\begin{eqnarray}
u &=& -(p_b-p_{+})^2 + (p_b - p_{-})^2, \nonumber\\ 
s &=& (p_{+} + p_{-})^2 , \nonumber \\
  u(s,m_s)&=&\sqrt{(s-(m_b + m_s )^2)
(s - (m_b -m_s )^2)}.
\label{kinvar}  
\end{eqnarray}
For subsequent use, we note that
$p_\pm = (E_\pm, \mbox{\boldmath $p_\pm$})$, and
 $q_\mu = (p_+ + p_-)_\mu$ is the
momentum transfer to the lepton-anti-lepton pair (hence $q^2=s$).
 We also define the
four-velocity of the $b$ quark, $ v_\mu = (p_b)_\mu/m_b$,
which we shall take subsequently to
be the same as that of the $B$ hadron, $v_\mu = (p_B)_\mu/M_B$.
Finally, we introduce the scaled variables $\hat{s}$ and $\hat{u}$
\begin{eqnarray}
\hat{s} &=& \frac{s}{m_b^2} ,\nonumber\\
\hat{u} &=& \frac{u}{m_b^2} = 2 v \cdot (\hat{p}_{+}-\hat{p}_{-}),
\label{kinvar2}
\end{eqnarray}
which in the decay \bsll are bounded as follows,
\begin{eqnarray}
& &-  \u(\s,\ms)< \u < + \u(\s,\ms)\, ,  \nonumber\\
\u(\s,\ms)  & = &
                 \sqrt{\left[ \s - (1 + \ms)^2 \right]
                 \left[ \s - (1 - \ms)^2 \right] }\, , \nonumber\\
& & 4 \ml^2< \s < (1-\ms)^2\, .
\end{eqnarray}  
where $\hat{m_i}$ and $\hat{p_i}_{\mu}$ 
 are the scaled quark masses and scaled momenta, respectively,
 $\hat{m}_i=m_i/m_b, ~{\hat{p}}_{i}=p_{i}/m_b$.
\subsection{NLO-corrected amplitude for \bsll in the effective Hamiltonian
approach}
Next, the  explicit 
expressions for the matrix element and (partial) branching ratios in
the decays  \bsll are presented in terms of the Wilson coefficients of the 
effective 
Hamiltonian obtained by integrating out the top quark and the $W^\pm$ bosons,
\begin{equation}\label{heffbsll}
{\cal H}_{eff}(b \to s + X)
  = {\cal H}_{eff} (b \to s + \gamma) -\frac{4 G_F}{\sqrt{2}} V_{ts}^* V_{tb}
\left[ C_9  O_9 +C_{10} O_{10} \right],
\end{equation}
where 
\begin{equation}\label{heffbsg}
{\cal H}_{eff}(b \to s +\gamma) = - \frac{4 G_F}{\sqrt{2}} V_{ts}^* V_{tb}
        \sum_{i=1}^{8} C_i (\mu)  O_i (\mu) \, .
\end{equation}
 Here, $V_{ij}$ are the CKM matrix elements and the
CKM unitarity has been used in factoring out the product $V_{ts}^\ast
V_{tb}$. The operator basis is chosen to be (here $\mu$ and $\nu$ are
Lorentz indices and $\alpha$ and $\beta$ are colour indices)
\begin{eqnarray}
 O_1 &=& (\bar{s}_{L \alpha} \gamma_\mu b_{L \alpha})
               (\bar{c}_{L \beta} \gamma^\mu c_{L \beta}),    \\
 O_2 &=& (\bar{s}_{L \alpha} \gamma_\mu b_{L \beta})
               (\bar{c}_{L \beta} \gamma^\mu c_{L \alpha}),    \\
 O_3 &=& (\bar{s}_{L \alpha} \gamma_\mu b_{L \alpha})
               \sum_{q=u,d,s,c,b}
               (\bar{q}_{L \beta} \gamma^\mu q_{L \beta}),    \\
 O_4 &=& (\bar{s}_{L \alpha} \gamma_\mu b_{L \beta})
                \sum_{q=u,d,s,c,b}
               (\bar{q}_{L \beta} \gamma^\mu q_{L \alpha}),    \\
 O_5 &=& (\bar{s}_{L \alpha} \gamma_\mu b_{L \alpha})
               \sum_{q=u,d,s,c,b}
               (\bar{q}_{R \beta} \gamma^\mu q_{R \beta}),    \\
 O_6 &=& (\bar{s}_{L \alpha} \gamma_\mu b_{L \beta})
                \sum_{q=u,d,s,c,b}
               (\bar{q}_{R \beta} \gamma^\mu q_{R \alpha}),    \\  
 O_7 &=& \frac{e}{16 \pi^2}
          \bar{s}_{\alpha} \sigma_{\mu \nu} (m_b R + m_s L) b_{\alpha}
                F^{\mu \nu},                                   \\
 O_8 &=& \frac{g}{16 \pi^2}
    \bar{s}_{\alpha} T_{\alpha \beta}^a \sigma_{\mu \nu} (m_b R + m_s L)  
          b_{\beta} G^{a \mu \nu},  \\
\end{eqnarray}
where $L$ and $R$ denote chiral projections, $L(R)=1/2(1\mp \gamma_5)$,
and the two additional operators involving the dileptons are:
\begin{eqnarray}
 O_9 &=& \frac{e^2}{16 \pi^2} \bar{s}_\alpha \gamma^{\mu} L b_\alpha
\bar{\ell} \gamma_{\mu} \ell , \nonumber\\
 O_{10} &=& \frac{e^2}{16 \pi^2} \bar{s}_\alpha \gamma^{\mu} L
b_\alpha \bar{\ell} \gamma_{\mu}\gamma_5 \ell ~.
\end{eqnarray}
The Wilson coefficients are given in the
literature (see, for example, \cite{Misiak1,buras}).

 With the help of the
effective Hamiltonian in eq.~(\ref{heffbsll})
 the matrix element for the decay \bsll can be written as, 
\begin{eqnarray}
	{\cal M (\mbox{\bsll})} & = & 
	\frac{G_F \alpha}{\sqrt{2} \pi} \, V_{ts}^\ast V_{tb} \, 
	\left[ \left( C_9^{\mbox{eff}} - C_{10} \right) 
		\left( \bar{s} \, \gamma_\mu \, L \, b \right)
		\left( \bar{l} \, \gamma^\mu \, L \, l \right) 
                \right. \nonumber \\
        & & \left.
                \; \; \; \; \; \; \; \; \;
                \; \; \; \; \; \; \; \; \;
		+ \left( C_9^{\mbox{eff}} + C_{10} \right) 
		\left( \bar{s} \, \gamma_\mu \, L \, b \right)
		\left( \bar{l} \, \gamma^\mu \, R \, l \right)  
		\right. \nonumber \\
	& & \left. 
                \; \; \; \; \; \; \; \; \;
                \; \; \; \; \; \; \; \; \;
		- 2 C_7^{\mbox{eff}} \left( \bar{s} \, i \, \sigma_{\mu \nu} \, 
		\frac{q^\nu}{q^2} (m_s L + m_b R) \, b \right)  
		\left( \bar{l} \, \gamma^\mu \, l \right) 
		\right] \, .
	\label{eqn:hamiltonian}
\end{eqnarray}
We have kept the $s$-quark mass term in the matrix element explicitly
and this will be kept consistently in the calculation of power corrections
and phase space. The above matrix element can be written in a compact form,
\begin{equation}
        {\cal M (\mbox{\bsll})} =
	\frac{G_F \alpha}{\sqrt{2} \pi} \, V_{ts}^\ast V_{tb} \, 
	\left( {\Gamma^L}_\mu \, {L^L}^\mu 
	+  {\Gamma^R}_\mu \, {L^R}^\mu \right) \, ,
\end{equation}
with
\begin{eqnarray}
	{L^{L/R}}_\mu & \equiv & 
		\bar{l} \, \gamma_\mu \, L(R) \, l \, , \\
	{\Gamma^{L/R}}_\mu & \equiv & 
		\bar{s} \left[ 
		R \, \gamma_\mu 
			\left( C_9^{\mbox{eff}} \mp C_{10} + 2 C_7^{\mbox{eff}} \, 
			\frac{\hat{\not{q}}}{\s} \right)
		+ 2 \hat{m}_s \, C_7^{\mbox{eff}} \, \gamma_\mu \, 
			\frac{\hat{\not{q}}}{\s} L 
		\right] b \, . 	
	\label{eqn:gammai}
\end{eqnarray}
 We recall that the
coefficient $C_9$ in LO is scheme-dependent. However, this is compensated
by an additional scheme-dependent part in the
(one loop) matrix element of $ O_9$ \cite{Misiak1,buras}. We call the
sum  $C_9^{\mbox{eff}}$, which is scheme-independent and enters in the
physical decay amplitude given above, with
\begin{equation}
C_9^{\mbox{eff}} (\hat{s}) \equiv C_9\eta({\hat{s}}) + Y(\hat{s}).
\end{equation}
The function $Y(\hat{s})$ is the one-loop matrix element of $ O_9$
 and is defined as \cite{misiak,buras}:
 \begin{eqnarray}
        Y(\s) & = & g(\mc,\s)
                \left(3 \, C_1 + C_2 + 3 \, C_3
                + C_4 + 3 \, C_5 + C_6 \right)
\nonumber \\
        & & - \frac{1}{2} g(1,\s)
                \left( 4 \, C_3 + 4 \, C_4 + 3 \,
                C_5 + C_6 \right) \nonumber \\
        & & - \frac{1}{2} g(0,\s) \left( C_3 +   
                3 \, C_4 \right) \nonumber \\
        & &     + \frac{2}{9} \left( 3 \, C_3 + C_4 +
                3 \, C_5 + C_6 \right) \nonumber \\
        & &     - \xi \, \frac{4}{9} \left( 3 \, C_1 +
                C_2 - C_3 - 3 \, C_4 \right),
                \label{eqn:y} \\
        \eta(\s) & = & 1 + \frac{\alpha_s(\mu)}{\pi}
                \omega(\s) ~.
\end{eqnarray}
Here, $\xi$ is dependent on the
dimensional regularization scheme \cite{Misiak1,buras}, with
 \begin{equation}
        \xi = \left\{
                \begin{array}{ll}
                        0       & \mbox{(NDR)}, \\
                        -1      & \mbox{(HV)},
                \end{array}
                \right.
\end{equation}
in the naive dimensional regularization (NDR) and the 't Hooft-Veltman (HV)
schemes.
 The function $\omega(\hat{s})$ represents the
$O(\alpha_s)$ correction from the one-gluon exchange in the matrix element of
$ O_9$ \cite{JK89}:
\begin{eqnarray}
\omega(\hat{s}) &=& -\frac{2}{9}\pi^2 -\frac{4}{3}{\mbox Li}_2(\hat{s})
-\frac{2}{3}
\ln \hat{s} \ln(1-\hat{s}) -
\frac{5+4\hat{s}}{3(1+2\hat{s})}\ln(1-\hat{s})\nonumber\\
&-& \frac{2\hat{s}(1+\hat{s})(1-2\hat{s})}{3(1-\hat{s})^2(1+2\hat{s})}
\ln \hat{s} + \frac{5 + 9\hat{s} -6\hat{s}^2}{6(1-\hat{s})(1+2 \hat{s})}~.
\label{omegahats}
\end{eqnarray}
                
 The function $g(z,\hat{s})$ includes the
charm quark-antiquark pair contribution
\cite{Misiak1,buras}:
 \begin{eqnarray}
g(z,\hat{s}) &=& -\frac{8}{9}\ln (\frac{m_b}{\mu})
 -\frac{8}{9} \ln z + \frac{8}{27} +\frac{4}{9}y
-\frac{2}{9}(2 + y) \sqrt{\vert 1-y \vert}\nonumber\\
&\times & \left[\Theta(1-y)(\ln\frac{1+\sqrt{1-y}}{1-\sqrt{1-y}} -i\pi )
+\Theta(y-1) 2 \arctan \frac{1}{\sqrt{y-1}} \right] ~,
\end{eqnarray}
\begin{equation}
g(0,\hat{s}) = \frac{8}{27}-\frac{8}{9}\ln (\frac{m_b}{\mu})
              -\frac{4}{9}\ln \hat{s} + \frac{4}{9}i\pi ~,
\end{equation}
where $y=4z^2/\hat{s}$.
With the help of the above expressions, the differential
decay width becomes, 
\begin{equation}
	{\rm d} \Gamma = \frac{1}{2 M_b} 
		\frac{{G_F}^2 \, \alpha^2}{2 \pi^2} 
		\left| V_{ts}^\ast V_{tb} \right|^2 
		\frac{{\rm d}^3 \mbox{\boldmath $p_+$}}{(2 \pi)^3 2 E_+} 
		\frac{{\rm d}^3 \mbox{\boldmath $p_-$}}{(2 \pi)^3 2 E_-} 
		\left( {W^L}_{\mu \nu} \, {L^L}^{\mu \nu} 
		+  {W^R}_{\mu \nu} \, {L^R}^{\mu \nu} \right) \, ,
\end{equation}
where $W_{\mu \nu}$ and $L_{\mu \nu}$ are the
hadronic and leptonic tensors, respectively.
The hadronic tensor $W_{\mu\nu}^{L/R}$
is related to the discontinuity in the forward scattering amplitude, 
denoted by
$T_{\mu \nu}^{L/R}$, through the relation $W_{\mu \nu} = 2 \, {\rm Im} \, 
T_{\mu \nu}$.  Transforming the integration variables 
 to $\hat{s}$, $\hat{u}$ and $v \cdot \hat{q}$, one can express the 
Dalitz distribution in \bsll (neglecting the lepton masses) as:  
\begin{equation}
	\frac{{\rm d} \Gamma}{{\rm d}\u \, {\rm d}\s} = 
		\frac{1}{2 \, m_b}
		\frac{{G_F}^2 \, \alpha^2}{2 \, \pi^2} 
		\frac{{m_b}^4}{256 \, \pi^4}
		\left| V_{ts}^\ast V_{tb} \right|^2 
		\, 2 \, {\rm Im} 
		\int {\rm d}(\z) \,
                \left( {T^L}_{\mu \nu} \, {L^L}^{\mu \nu}
                +  {T^R}_{\mu \nu} \, {L^R}^{\mu \nu} \right) \, ,
	\label{eqn:dgds}
\end{equation}
with
\begin{eqnarray}
	{T^{L/R}}_{\mu \nu} & \equiv & 
	i \, \int {\rm d}^4 y \, e^{-i \, \hat{q} \cdot y}	
	\left< B \left| {\rm T} \left\{ 
		{{\Gamma_1}^{L/R}_\mu} (y), 
		{\Gamma_2}^{L/R}_\nu (0) \right\} \right| B \right>\, , \\
	{L^{L/R}}^{\mu \nu} & \equiv & 
	\sum_{spin} 
	\left[ \bar{v}^{L/R}(p_+) \, \gamma^\mu \, u^{L/R}(p_-) \right]
	\left[ \bar{u}^{L/R}(p_-) \, \gamma^\nu \, v^{L/R}(p_+) \right]
		\nonumber \\
	& = & 2 \left[ {p_+}^\mu \, {p_-}^\nu + {p_-}^\mu \, {p_+}^\nu 
		- g^{\mu \nu} (p_+ \cdot p_-) 
		\mp i \epsilon^{\mu \nu \alpha \beta} \, 
			{p_+}_\alpha \, {p_-}_\beta \right] \, , 
\end{eqnarray}
where ${{\Gamma_1}^{L/R}_\mu}^\dagger = {\Gamma_2}^{L/R}_\mu = 
\Gamma^{L/R}_\mu $, given in eq. (\ref{eqn:gammai}). The Dalitz distribution 
eq.~(\ref{eqn:dgds}) contains the explicit $O(\alpha_s)$-improvement, and 
the two
distributions in which we are principally interested in can be obtained
by straight-forward integrations.
\subsection{Leading power $(1/m_b)$ corrections in the decay  \bxsll}
The next task is to expand the forward scattering amplitude 
$T_{\mu \nu}$ in the inverse powers in  
 $1/{m_b}$. Suppressing the Lorentz indices for the time being, this
expansion can be formally represented as:  
\begin{eqnarray}
	\int {\rm d}^4 y \, e^{-i \, \hat{q} \cdot y} 
	\left< B \left| {\rm T} \left\{ \Gamma_1 (y), 
	\Gamma_2 (0) \right\} \right| B \right> 
	& = & - \frac{1}{m_b} \left[ 
	\left< B \left| {\cal O}_0 \right| B \right> 
	+ \frac{1}{2 \, m_b} \left< B \left| {\cal O}_1\right| B \right> 
	\right. 
	\nonumber \\
	& & \left.
	\, \, \, \, \, \, \, 	
	+ \frac{1}{4 \, {m_b}^2} \left< B \left| {\cal O}_2 \right| B \right>  
	+ \cdots \right] \, ,
\label{toproduct}
\end{eqnarray}
and the expressions for the operators ${\cal O}_0, {\cal O}_1$ and ${\cal 
O}_2$ 
are given explicitly  in \cite{falketal}. They are obtained by expanding the
propagators in the Feynman diagrams contributing to the time-ordered product
on the l.h.s. of the above equation (see Fig. 1 in \cite{falketal}),
using $p_b{_\mu}=m_b v_\mu + k_\mu$, 
fixing the four-velocity of the external $b$ quark field to be $v_\mu$
and treating the components of the ``residual momentum" $k_\mu$
to be much smaller than $m_b$.

 As is well known, the leading power corrections can be
parametrized in terms of the matrix elements of the kinetic energy and
magnetic moment operators, called $\lambda_1$ and $\lambda_2$,
respectively, and defined as,
 \begin{eqnarray}
        \left< B \left| \bar{h} \, (i \, D)^2 \, h \right| B\right>   
                & \equiv & 2 \, M_B \, \loo
                \, , \nonumber \\
        \left< B \left| \bar{h} \, \frac{-i}{2} \sigma^{\mu \nu}
                \, G_{\mu \nu} \, h \right| B \right>
                & \equiv & 6 \, M_B \, \lto
                \, ,
\label{hqetpar}
\end{eqnarray}
where $B$ denotes the pseudoscalar $B$ meson, $D_\mu$ is the covariant   
derivative and $G_{\mu \nu}$ is the QCD field strength tensor.
The two-component effective field in the HQE approach $h(y)$ is
related to the QCD field $b(y)$ through the expansion,
\begin{equation}
\label{bhrel}
b(y) = e ^{i m_b v.y} \left[1+i \frac{\not{D}}{2 m_b} + ...\right] 
h(y),   
\label{btohx}
\end{equation}
where $\not{D} = D_\mu \gamma^\mu$.
 The parameters $\loo$ and $\lto$ are related through the
quantity $\bar{\Lambda}$ to the hadron masses \cite{Luke90},
\begin{eqnarray} 
m_B &=& m_b + \bar{\Lambda} - \frac{\loo + 3\lto}{2m_b}+..., \nonumber\\
m_{B^*} &=& m_b + \bar{\Lambda} - \frac{\loo -\lto}{2m_b}+....
\end{eqnarray}
>From the $B - B^{*}$ mass difference, one obtains $\lambda_2 \simeq 0.12$
GeV$^2$. The quantity $\loo$  has been determined from
QCD sum rules \cite{kineticsr,neubertsr96} and data \cite{kineticdata}. Its
present value is subject to a certain theoretical dispersion,
estimated somewhere between $\lambda_1=-(0.52 \pm 0.12)$ GeV$^2$ (Ball and
Braun in \cite{kineticsr}) to $\lambda_1=-(0.10 \pm 0.05)$ GeV$^2$
(Neubert \cite{neubertsr96}).

 Concerning the definitions of the operators in eq.~(\ref{toproduct}), we 
follow the
prescription given in \cite{falketal}, in which the leading operator 
${\cal O}_0$ is defined in terms of the ``full" four-component field $b(y)$,
\begin{equation}
{\cal O}_0(y) = \frac{1}{x} \bar{b} \Gamma_1 (\not{v} -\not{\q} + 
\ms) \Gamma_2 b ,
\end{equation}
where $x \equiv 1 + \s - 2 \, (\z) - \ms^2 + i \, \epsilon $.
The other two subleading operators ${\cal O}_1$ and ${\cal O}_2$ are, however,
written in terms of the two-component effective fields $h(y)$, which is
related to the field $b(y)$ through the expansion given in eq.~(\ref{btohx}).
 Of these, the
expression for ${\cal O}_2$ involving the expansion of the one-gluon 
graph is obtained by a non-trivial derivation, which we have checked, and 
it agrees with the one given in eq.~(3.8) of \cite{falketal}.
(Likewise, we agree with the expression for ${\cal O}_1$ given in
eq.~(3.6) of \cite{falketal}.) For the sake of completeness, we give below
the explicit expression for ${\cal O}_1$ and  ${\cal O}_2$ 
given in \cite{falketal}: 
\begin{equation}
{\cal O}_1(y) = \frac{2}{x} \bar{h} \Gamma_1 \gamma^{\alpha} \Gamma_2 i
D_{\alpha} h 
- \frac{4}{x^2}(v-\q)^{\alpha} \bar{h} \Gamma_1 (\not{v} -\not{\q} + 
\ms) \Gamma_2 i D_{\alpha} h ,
\end{equation}
and
\begin{eqnarray}
{\cal O}_2(y) & = & \frac{16}{x^3} (v- \q)^{\alpha} (v- \q)^{\beta} \bar{h}
\Gamma_1 (\not{v} -\not{\q} + \ms) \Gamma_2 i D_{\alpha} i D_{\beta} h
- \frac{4}{x^2} \bar{h} \Gamma_1 (\not{v} -\not{\q} + \ms) \Gamma_2 (i D)^2 h
\nonumber \\
& - & \frac{4}{x^2}(v-\q)^{\beta} \bar{h} \Gamma_1 \gamma^{\alpha} \Gamma_2 
(i D_{\alpha} i D_{\beta}+ i D_{\beta} i D_{\alpha}) h + \frac{2}{x^2} \ms
 \bar{h} \Gamma_1 i \sigma_{\alpha \beta} \Gamma_2 G^{\alpha \beta} h
 \\
&+& \frac{2}{x^2} i \epsilon^{\mu \lambda \alpha \beta}
 ( v- \q)_{\lambda}\bar{h}
\Gamma_1 \gamma_{\mu} \gamma_{5} \Gamma_2 G_{\alpha \beta} h+
\frac{2}{x} \bar{h} (\gamma^{\beta} \Gamma_1 \gamma^{\alpha} \Gamma_2+
 \Gamma_1 \gamma^{\beta} \Gamma_2 \gamma^{\alpha})  i D_{\beta} i D_{\alpha} h
\nonumber \\
&-& \frac{4}{x^2}(v-\q)^{\alpha} \bar{h} \gamma^{\beta}
 \Gamma_1  (\not{v} -\not{\q} + \ms) \Gamma_2 i D_{\beta} i D_{\alpha} h-
\frac{4}{x^2}(v-\q)^{\alpha} \bar{h} \Gamma_1 
(\not{v} -\not{\q} + \ms) \Gamma_2 \gamma^{\beta} i D_{\alpha} i 
D_{\beta} h . \nonumber 
\end{eqnarray}
 Using Lorentz decomposition, the tensor $T_{\mu \nu}$ can be expanded in 
terms of three structure functions, 
\begin{equation}
	T_{\mu \nu} = -T_1 \, g_{\mu \nu} + T_2 \, v_\mu \, v_\nu 
		+ T_3 \, i \epsilon_{\mu \nu \alpha \beta} \, 
			v^\alpha \, \hat{q}^\beta \, ,
\end{equation}
where the structure functions which do not contribute to the
amplitude in the limit of massless leptons have been neglected.
After contracting the hadronic and leptonic tensors, one finds
\begin{equation}
	{T^{L/R}}_{\mu \nu} \, {L^{L/R}}^{\mu \nu} = 
		{m_b}^2 \left\{ 2 \, \s \, {T_1}^{L/R} 
		+ \left[ (\z)^2 - \frac{1}{4} \u^2 - \s \right] {T_2}^{L/R} 
		\mp \s \, \u \, {T_3}^{L/R} \right\} \, . 
	\label{eqn:tlr}
\end{equation}
We remark here that the $T_3$ term will contribute to 
the FB asymmetry but not to the branching ratio or
the dilepton invariant mass spectrum in the decay \bxsll.

 The results of the power corrections
to the structure functions $T_i$ can be decomposed into the sum of
various terms, denoted by $T_{i}^{(j)}$, which can be traced back to well 
defined pieces in the evaluation of the time-ordered product given above:
\begin{equation}
T_{i}(v.\hat{q},\s) = \sum_{j=0,1,2,s,g,\delta} T_{i}^{(j)}(v.\hat{q}, \s)\,.
\label{Tijhqet}
\end{equation}
The expressions for $T_{i}^{(j)}(v.\hat{q},\s)$ calculated up to 
$O(M_B/m_b^3)$ are given in Appendix A.
They contain the parton model expressions $T_{i}^{(0)}(v.\hat{q},\s)$ and 
the power 
correction in the HQE approach which depend on the two HQET-specific 
parameters $\loo$ and $\lto$  defined in eqs.~(\ref{hqetpar}).
Note that the $s$-quark mass terms are explicitly kept in
 $T_{i}^{(j)}(v.\hat{q},\s)$.

\par 
 From the expressions for $T_{i}^{(j)}$ given in Appendix A,
 we see that $T_{i}^{(0)} (i=1,2,3)$ are of order 
$M_B/(m_b)$
 and the rest $T_{i}^{(1)}$ , $T_{i}^{(\delta)}, T_{i}^{(2)}, T_{i}^{(s)}$ 
and  $ T_{i}^{(g)}$ are all of order ${M_B}\lambda_1/{ m_ {b} }^3$
or ${M_B}\lambda_2/{ m_ {b} }^3$.
Since the ratio $M_B/m_b = 1 + O(1/m_b)$, we note that the Dalitz 
distribution in \bxsll has linear corrections in $1/m_b$.
 The origin  of the 
various terms in the expansion given in eq.~(\ref{Tijhqet}) is as follows:
\begin{itemize}
\item The contributions to
 $T_{i}^{(1)}$ come from the matrix element of those terms in the operator 
${\cal O}_2$, which
 originate from expanding the spinor of the heavy quark field $b(x)$ in terms
of the spinor of the heavy quark effective theory $h(x)$.
\item  The remaining contributions from the matrix element of the operator 
${\cal O}_2$ are denoted by $T_{i}^{(2)}$
 and $ T_{i}^{(g)}$, with
 $ T_{i}^{(g)}$ originating from the matrix element of the one gluon
 emission diagram and the rest being $T_{i}^{(2)}$.
\item  The contributions  
denoted by $T_{i}^{(\delta)}$ arise from the matrix element of the 
operator ${\cal O}_1$.
 In the leading order in $(1/m_b)$ this matrix element vanishes, but in the
sub-leading order it receives a non-trivial contribution which can be 
calculated by using the equation of motion.
\item  Finally, the contributions 
$T_{i}^{(s)}$ arise from the matrix element of the scalar 
operator $\bar{b} b$.
\end{itemize}
Concerning the last point noted above, we recall that
 the scalar current can be written in terms of 
the vector current plus higher dimensional operators as \cite{georgi}
\begin{equation}
\bar{b} b = v_\mu \bar{b} \gamma^\mu b + \frac{1}{2m_b^2}
\bar{h} \left[ (iD)^2 - (v.iD)^2 + s^{\mu \nu} G_{\mu\nu} \right] h +...,
\end{equation}
with $G_{\mu \nu}=
[iD_\mu , i D_\nu]$ and $s^{\mu \nu} = (-i/2) \sigma^{\mu \nu}$.
We note that in deriving $T_{i}^{(s)}$, use has been made of  
the conservation of the $b$-number current in QCD, which yields the 
normalization:
\begin{equation}
\langle B \vert \bar{b} \gamma_\mu b \vert B \rangle = 2 (p_B)_{\mu} .
\end{equation}

Finally, after doing the integration on the complex plane 
$\z$ (see Fig.~1 in \cite{manoharwise} for the analytic structure of
$T_{\mu\nu}$ and the contour of integration), we derive the 
double differential branching ratio in \bxsll. The
result can be expressed as,
\begin{eqnarray}
\frac{{\rm d}{\cal B}}{{\rm d}\s \, {\rm d}\u} & = & 
		{\cal B}_0 \, 
	\left( \left\{ 
	\left[ \left( 1 - \ms^2 \right)^2 - \s^2 - \u^2 
	- \frac{1}{3} \left( 2 \, \lo (-1 + 2 \ms^2 -\ms^4 - 2 \, \s + \s^2) 
\right. \right. \right. \right.
\nonumber \\
& &  \left. \left. \left. \left.
		+ 3 \, \lt (-1 +6 \ms^2 -5 \ms^4 - 8 \, \s + 5 \, \s^2) 
	\right)
	\right] \left( |C_9^{\mbox{eff}}|^2 + |C_{10}|^2 \right)
	\right. \right.
	\nonumber \\
	& & \left. \left.
	+ \, \left[ 4 \left( 
   	1 - \ms^2 -\ms^4+\ms^6-8\ms^2 \s -\s^2 -\ms^2 \s^2+ \u^2 +\ms^2 \u^2 
	      \right)  
	\right. \right.  \right. 
	\nonumber \\
	& & 
	- \frac{4}{3} \left( 2 \, \lo (-1 + \ms^2 +\ms^4 -\ms^6 + 2 \, \s +10 \ms^2 \s + \s^2 +\ms^2 \s^2)
\right.
\nonumber \\
& & \left. \left.
		+ 3 \, \lt (3 +5 \ms^2 -3 \ms^4 -5 \ms^6 + 4 \, \s + 28 \ms^2 \s + 5 \, \s^2 + 5 \ms^2 \s^2) \right)
	\right] \frac{|C_7^{\mbox{eff}}|^2}{\s}
	\nonumber \\
	& &  
	- 8 \, \left[ \left( \s (1 + \ms^2) - (1 - \ms^2)^2 \right)
	+ \frac{2}{3}  \lo (-1 +2 \ms^2 - \ms^4 + \s + \ms^2 \s) 
 \right.
\nonumber \\
& & \left.
		+  \lt (5 \ms^2 - 5 \ms^4 + 2 \s + 5 \ms^2 \s) 
	\right] Re(C_9^{\mbox{eff}}) \, C_7^{\mbox{eff}}
	\nonumber \\
	& & \left. \left.
	+ 2 \, \left[ 2 +  \lo + 5 \, \lt  \right] 
		\u \, \s \, Re(C_9^{\mbox{eff}}) \, C_{10}
	\right. \right.
	\nonumber \\
	& & \left. \left.
	+  4 \, \left[ 2 \left( 1 + \ms^2 \right)  
	+  \lo (1+ \ms^2)+  \lt (3+ 5 \ms^2)  \right] \u \, Re(C_{10}) \, C_7^{\mbox{eff}}
	\right\} \theta \left[ {\u(\s,\ms)}^2 - \u^2 \right]
	\right.
	\nonumber \\
	& & \left. 
	- E_1 (\s, \u) \, \delta \left[ {\u(\s,\ms)}^2 - \u^2 \right] 
	- E_2 (\s, \u) \, \delta^\prime \left[ {\u(\s,\ms)}^2 - \u^2 \right] 
	\right) \, , 
	\label{eqn:dddw}
\end{eqnarray}
where $\lo=\lambda_{1}/m_{b}^2$ and $\lt=\lambda_{2}/m_{b}^2$.
The auxiliary functions $E_i (\s, \u)$ ($i = 1,2$), introduced  here
for ease of writing, are given explicitly in Appendix B.
The boundary of the Dalitz distribution is as usual determined by the
argument of the $\theta$-function and in the $(\u,\s)$-plane
it has been specified earlier.
 The analytic 
form of the result (\ref{eqn:dddw}) is very similar to the corresponding 
double differential
distributions derived by Manohar and Wise in \cite{manoharwise} for
the semileptonic decays $B \to (X_c,X_u) \ell \nu_\ell$. Further comparisons
with this work in the $V-A$ limit for the single differential  and 
integrated rates are given in Appendix C.  
  
It has become customary to express the branching ratio for \bxsll in terms
of the well-measured semileptonic branching ratio ${\cal B}_{sl}$
for the decays $B \to (X_c,X_u) \ell \nu_\ell$. This fixes
the normalization constant ${\cal B}_0$ to be,
\begin{equation}
        {\cal B}_0 \equiv
                {\cal B}_{sl} \frac{3 \, \alpha^2}{16 \pi^2} \frac{
    {\vert V_{ts}^* V_{tb}\vert}^2}{\absvcb^2} \frac{1}{f(\mc) \kappa(\mc)}
                \, ,
\label{eqn:seminorm}
\end{equation}
where
\begin{equation}
        f(\mc) = 1 - 8 \, \mc^2 + 8 \, \mc^6 - \mc^8 - 24 \, \mc^4 \, \ln \mc
        \label{eqn:fr}
\end{equation}
is the phase space function for $\Gamma (B \rightarrow X_c l \nu)$
in the lowest order (i.e., parton model) and
the function $\kappa(\mc)$ accounts for both the $O(\alpha_s)$ QCD 
correction to 
the semi-leptonic decay  width \cite{aliqcd,JK89} and the leading order
$(1/m_b)^2$ power correction \cite{georgi}. Written explicitly, it reads as:
\begin{equation}
\kappa(\mc) = 1 - \frac{2 \alpha_s(m_b)}{3 \pi} g (\mc)
    + \frac{h(\mc)}{2 m_b^2},
\end{equation}
where the two functions are:
\begin{eqnarray}
g(\mc) &=& (\pi^2-\frac{31}{4})(1-\mc)^2 + \frac{3}{2} \, , \nonumber\\
h(\mc) &=& \lambda_1 + \frac{\lambda_2}{f(\mc)} \left[ -9 +24 \mc^2
-72\mc^4 + 72\mc^6 -15\mc^8 -72 \mc^4 \ln \mc \right]\, .
\label{eqn:ghr}
\end{eqnarray}

Finally, after integrating over the variable $\hat{u}$, we derive the 
differential branching ratio in the scaled dilepton invariant mass 
for \bxsll , 
\begin{eqnarray}
	\frac{{\rm d}{\cal B}}{{\rm d}\s} & = & 2 \; {\cal B}_0 
		\left\{ 
                  \left[
   		\frac{2}{3} \u(\s,\ms) ((1 - \ms^2)^2 + \s (1+\ms^2) -2 \s^2) 
   	+	\frac{1}{3} (1 -4 \ms^2 + 6 \ms^4 -4 \ms^6 + \ms^8 -\s 
\right. \right.
		\nonumber \\
	& &   
+ \ms^2 \s +
\ms^4 \s - \ms^6 \s -3 \s^2 -2 \ms^2 \s^2 -3 \ms^4 \s^2 + 5 \s^3 +5 \ms^2 \s^3-2 \s^4 ) \frac{\lo}{ \u(\s,\ms)}  
\nonumber \\
  & &        		+ \left( 1 -8 \ms^2 + 18 \ms^4 -16 \ms^6 + 5 \ms^8 -\s 
-3 \ms^2 \s + 9 \ms^4 \s -5 \ms^6 \s -15 \s^2 -18 \ms^2 \s^2 
\right.
\nonumber \\
& &
\left. \left.
-15 \ms^4 \s^2 + 25 \s^3
 + 25 \ms^2 \s^3 -10 \s^4 \right) \frac{\lt}{ \u(\s,\ms)}  
			\right] 
     		\left( |C_9^{\mbox{eff}}|^2 + |C_{10}|^2 \right)
		\nonumber \\
	&  &
            + 	\left[
	 	 \frac{8}{3} \u(\s, \ms) (2 (1+\ms^2)(1-\ms^2)^2-
                (1+14 \ms^2 +\ms^4) \s -(1+\ms^2) \s^2 )
		\right.
		\nonumber \\
  	& &  
          +	\frac{4}{3} (2 - 6 \ms^2 + 4 \ms^4 +4 \ms^6 -6 \ms^8+
 2 \ms^{10} -5 \s -12 \ms^2 \s + 34 \ms^4 \s -12 \ms^6 \s -5 \ms^8 \s + 3 \s^2
\nonumber \\
& & 
 + 29 \ms^2 \s^2 + 29 \ms^4 \s^2 
+3 \ms^6 \s^2+ \s^3 -10 \ms^2 \s^3 +\ms^4 \s^3-\s^4-\ms^2 \s^4)
  \frac{\lo}{ \u(\s,\ms)} + 4  \left(-6 + 2 \ms^2
\right.
\nonumber \\
& &
 + 20 \ms^4 -12 \ms^6 - 14 \ms^8 +10 \ms^{10} +   3 \s+   16 \ms^2 \s +
 62 \ms^4 \s  -56 \ms^6 \s -25 \ms^8 \s + 3 \s^2
\nonumber \\
& & \left. \left.
  + 73 \ms^2 \s^2 + 101 \ms^4 \s^2 +15 \ms^6 \s^2+ 5 \s^3-26 \ms^2 \s^3+
5 \ms^4 \s^3 -5 \s^4-5 \ms^2 \s^4 \right) \frac{\lt}{ \u(\s,\ms)}
                \right] \frac{|C_7^{\mbox{eff}}|^2}{\s}
		\nonumber \\
	& &	
         +      \left[
		8 \u(\s, \ms) ((1-\ms^2)^2-(1+\ms^2) \s)
                + 4 ( 1 - 2 \ms^2 +\ms^4  - \s-\ms^2 \s) \; \u(\s,\ms) \;  \lo 
\right.
\nonumber \\
 & &
     + 4 \left( -5 +30\ms^4-40 \ms^6 +15 \ms^8 -\s + 21 \ms^2 \s +
 25 \ms^4 \s -45 \ms^6 \s+ 13 \s^2 + 22 \ms^2 \s^2\right.
\nonumber \\
& & 
\left. \left. \left.
+45 \ms^4 \s^2 -7 \s^3-15 \ms^2 \s^3 \right)\frac{\lt}{ \u(\s,\ms)}    
			\right] Re(C_9^{\mbox{eff}}) \, C_7^{\mbox{eff}} 
		\right\}
\, .
\end{eqnarray}
Another interesting quantity is the FB asymmetry
defined in \cite{amm91,agm94}
\begin{equation}
        \frac{{\rm d}{\cal A}(\s)}{{\rm d}\s} = \int_0^1
                \frac{{\rm d}^2 {\cal B}}{{\rm d}\s \, {\rm d}z} \, {\rm d}z
                - \int_{-1}^{0}
                \frac{{\rm d}^2 {\cal B}}{{\rm d}\s \, {\rm d}z} \, {\rm d}z
\label{eqn:fbasy}
        \, ,
\end{equation}
where $z \equiv \cos \theta$ is the angle of $\ell^+$ measured w.r.t. the
$b$-quark direction in the dilepton c.m. system.
 The leading power corrected expression for the
FB-asymmetry ${\cal A}(\s)$ is:
\begin{eqnarray}
        \frac{{\rm d}{\cal A}(\s)}{{\rm d}\s} & = &
                - 2 \; {\cal B}_0
                \left\{
                \left[
                2 (\u(\s ,\ms))^2 \s
        + \frac{\s}{3} (3 -6 \ms^2 + 3 \ms^4 +2 \s -6 \ms^2 \s+3 \s^2) \lo
\right. \right.
\nonumber \\
& & \left. 
 +  \s \, (-9 -6 \ms^2 + 15 \ms^4 -14 \s -30 \ms^2 \s+ 15 \s^2) \, \lt 
\right]  
                         \, Re(C_9^{\mbox{eff}}) \, C_{10}
                \nonumber \\
        & &
                + \left[ 4 (\u(\s ,\ms))^2 (1+\ms^2)
+ \frac{2}{3}\; (1+\ms^2) \;
 (3 -6 \ms^2 + 3 \ms^4 + 2 \s -6 \ms^2 \s + 3\s^2) \lo
\right. \nonumber\\
& & \left. \left.
                + 2 (-7 -3 \ms^2 -5 \ms^4
+15 \ms^6 -10 \s -24 \ms^2 \s-30 \ms^4 \s + 9 \s^2+ 15 \ms^2 \s^2)\,
 \lt \right]\,  Re(C_{10}) \, C_7^{\mbox{eff}}\,  \right\} .
\nonumber\\
&& 
\end{eqnarray}
>From the experimental point of view, a more useful quantity is the
normalized FB-asymmetry, obtained by normalizing $d{\cal A}/d\s$ with the 
dilepton mass distribution, $d{\cal B}/d\s$,
\begin{equation}
\frac{{\rm d} \overline{{\cal A}}}{{\rm d}\s} =  \frac{{\rm d}{\cal A}}{{\rm 
d}\s}/
 \frac{{\rm d}{\cal B}}{{\rm d}\s}
\, .   
\end{equation}
This asymmetry, which we recall is defined in the dilepton c.m.s.~frame, is 
identical to the
energy asymmetry introduced in \cite{choetal}, which is defined in the $B$ 
rest frame, as shown in Appendix D.

 The results derived for the $O(\alpha_s)$-improved and power-corrected 
 Dalitz distribution, dilepton invariant mass,
and FB-asymmetry in \bxsll are the 
principal new results in this section.
It is useful to write the 
corresponding expressions in the limit $m_s=0$. For the dilepton invariant
mass distribution, we get 
\begin{eqnarray}
	\frac{{\rm d}{\cal B}}{{\rm d}\s} & = & 2 \; {\cal B}_0 
		\left\{ 
                  \left[
   		\frac{1}{3} (1-\s)^2 (1+2 \s) \; (2 + \lo) 
      		+ ( 1 - 15  \s^2 + 10 \s^3\right) \lt
			\right] 
     		\left( |C_9^{\mbox{eff}} |^2 + |C_{10}|^2 \right)
		\nonumber \\
	& & 
             +	\left[
	 	 \frac{4}{3} (1-\s)^2 (2+ \s) \; (2 + \lo)
                + 4  \left( -6 -3  \s + 5 \s^3 \right) \lt
                \right] \frac{|C_7^{\mbox{eff}}|^2}{\s}
		\nonumber \\
	& & 	\left.
           +    \left[
		4 (1-\s)^2 (2+ \lo)
               + 4  \left( -5 -6  \s + 7 \s^2 \right) \lt   
				\right] Re(C_9^{\mbox{eff}}) \, C_7^{\mbox{eff}} 
		\right\}\, .
\label{eqn:dbds0}
\end{eqnarray}

 The (unnormalized) FB asymmetry reads as,
\begin{eqnarray}
	\frac{{\rm d}{\cal A}}{{\rm d}\s} & = & 
		- 2 \; {\cal B}_0 
                \left\{
		\left[ 
		2 (1 - \s)^2 \s 
		+ \frac{\s}{3} (3+2 \s +3 \s^2) \lo 
    		+  \s \, (-9 -14 \s + 15 \s^2) \, \lt \right] 
			 \, Re(C_9^{\mbox{eff}} ) \, C_{10}
		\right. 
		\nonumber \\
	& &	\left. 
		+ \left[ 4 (1-\s)^2   
 		+ \frac{2}{3} (3+ 2 \s + 3\s^2) \lo 
          	+ 2 (-7 -10 \s + 9 \s^2)\, \lt \right] \, Re(C_{10}) \, C_7^{\mbox{eff}} 
		 \,  \right\}\, .
\end{eqnarray}

 A direct comparison of our result for
the dilepton invariant mass distribution given in eq.~(\ref{eqn:dbds0}) 
 above can now be made with
the differential decay width ${\rm d}\Gamma(\mbox{\bxsll})/
{\rm d}\hat{s}$ derived in eq.~(3.21) of the paper by  FLS
\cite{falketal}. To that end, one has to
 take into account the (obvious) normalization difference between
the decay width and branching ratio, rewrite the quantities $A^{i}$ and
$B^{i}$ used in FLS \cite{falketal} in terms of the Wilson coefficients
$C_{7}^{\mbox{eff}}, C_9^{\mbox{eff}}$ and $C_{10}$ used by us, with 
$ A^{R/L} = C_{9}^{\mbox{eff}} \pm C_{10} $ and 
$B^{R/L}=-2 C_{7}^{\mbox{eff}}$, and drop the
explicit $O(\alpha_s)$-improvement in the coefficient $C_{9}^{\mbox{eff}}$, as
FLS did not include it in their calculations. The resulting expression is:
\begin{eqnarray}
	\frac{{\rm d}{\cal B_{FLS}}}{{\rm d}\s} & = & 4 \; {\cal B}_0 
		(1-\s) \left\{ 
                  \left[
   		\frac{1}{3} (1-\s) (1+2 \s)
\right. \right.
\nonumber \\
  & & \left. \left.
      +      \frac{1}{6} (5 + 3 \s -2 \s^2)\lo 
      		+\frac{1}{2} ( 1 + 15  \s - 10 \s^2\right) \lt
			\right] 
     		\left( |C_9^{\mbox{eff}} |^2 + |C_{10}|^2 \right)
		\nonumber \\
	& & 
             +	\left[
	 	 \frac{4}{3} (1-\s) (1 + \frac{2}{\s})-
                 \frac{2}{3} (1 +\s) \lo - 10 (1+\s) \lt
                \right] |C_7^{\mbox{eff}}|^2
		\nonumber \\
	& & 	\left.
           +    \left[
		4 (1-\s)-2 (-\frac{5}{3}+\s) \lo + 2 (5 -7 \s) \lt
	\right] Re(C_9^{\mbox{eff}}) \, C_7^{\mbox{eff}} 
		\right\}\, .
\label{eqn:dbds0fl}
\end{eqnarray}

\begin{figure}[htb]
\vskip -2.0truein
\centerline{\epsfysize=7in
{\epsffile{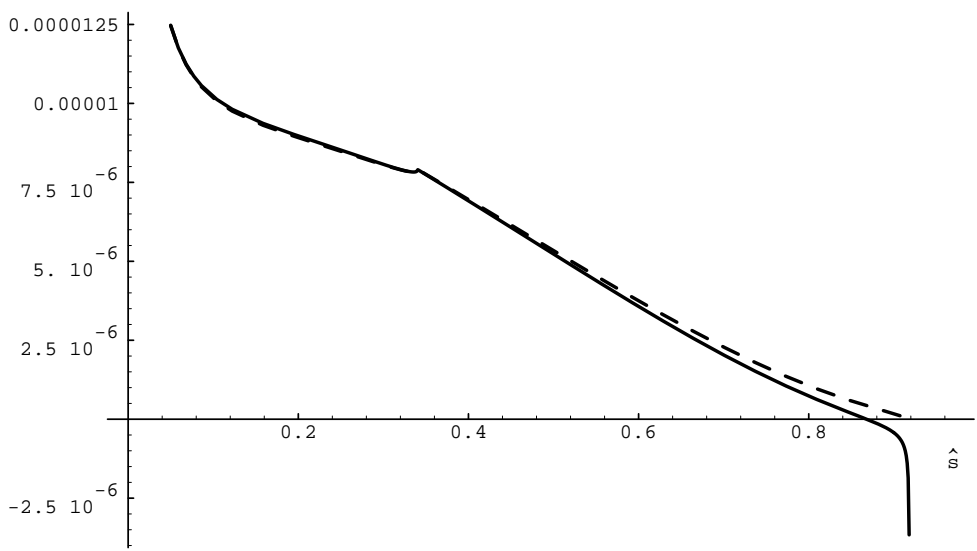}}}
\vskip -2.0truein
\caption[]{Dilepton invariant mass spectrum ${\rm d}{\cal B} (B \to X_s e^+
e^-)/{\rm d} \hat{s}$ in the parton model (dashed curve) and with leading 
power corrections calculated in the
HQE approach (solid curve). The parameters used are given in
Table \ref{parameters}.}
\label{fig:hillerfig1}
\end{figure}

\begin{figure}[htb]
\vskip -2.0truein
\centerline{\epsfysize=7in
{\epsffile{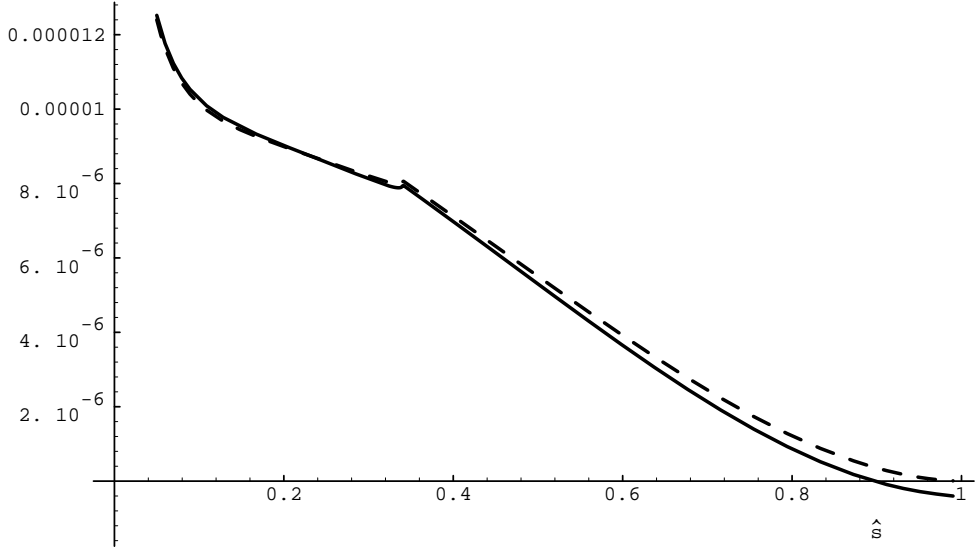}}}
\vskip -2.0truein
\caption[]{Dilepton invariant mass spectrum ${\rm d}{\cal B} (B \to X_s e^+ 
e^-)/{\rm d} \hat{s}$ with power corrections calculated in the
HQE approach. The solid curve corresponds to our calculation  and the
dashed curve results from eq. (3.21) of FLS \protect\cite{falketal} with
$m_s=0$. The other parameters are given in 
Table \ref{parameters}.}
\label{fig:flvsus}
\end{figure}

\begin{figure}[htb]
\vskip -2.0truein
\centerline{\epsfysize=7in
{\epsffile{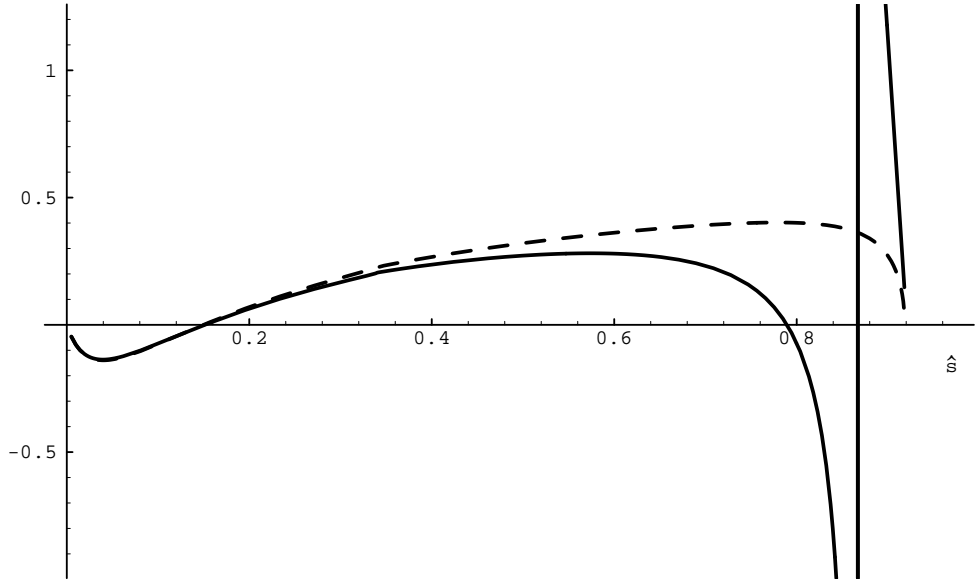}}}
\vskip -2.0truein   
\caption[]{FB asymmetry (normalized) ${\rm d}\overline{{\cal A}} (B \to 
X_s e^+
e^-)/{\rm d} \hat{s}$ in the parton model and with power corrections
calculated in the
HQE approach. The solid curve corresponds to the HQE spectrum and the
dashed curve is the parton model result. The parameters used are given in
Table \ref{parameters}.}
\label{fig:hillerfig2}
\end{figure}

We would like to make the following observations:
\begin{itemize}
\item The results derived here (eq. (52)) and in FLS \cite{falketal} (eq. 
(54)) reproduce the known
parton model expression for the dilepton invariant mass distribution in the 
limit $\lambda_1 \to 0$ and $\lambda_2 \to 0$.
\item The power corrections themselves, i.e. the expressions multiplying the 
constants 
$\lambda_1$ and $\lambda_2$, are {\it different} in the two derivations.
\item The power-corrected dilepton invariant mass
distribution derived by us retains the characteristic
$1/\hat{s}$ behaviour following from the one-photon exchange 
in the parton model, in contradiction to the observations 
made in \cite{falketal}. This difference can be seen by comparing the two
expressions multiplying the Wilson coefficient $|C_7^{\mbox{eff}}|^2$.
\item Leading order power corrections in the dilepton mass distribution are 
found to be small over a good part of the dilepton mass $\s$.
 However, we find
that the power corrections become increasingly
large and negative as one approaches $\hat{s} \to \hat{s}^{max}$.
 Since the parton model spectrum falls steeply near the
end-point $\hat{s} \to \hat{s}^{max}$, this leads to the
uncomfortable result that the power corrected
dilepton mass distribution becomes negative for the high dilepton masses - in
contradiction to the observations made in \cite{falketal}. We show in
Fig.~\ref{fig:hillerfig1} this distribution in the parton model and the
HQE approach, using the central values of the parameters in Table 
\ref{parameters}.
\item We note that the correction proportional to the kinetic energy
term $\hat{\lambda}_1$ renormalizes the parton model invariant 
mass distribution
multiplicatively by the factor $(1 + \lambda_1/(2m_b^2))$, i.e. no new 
functional dependence in $\hat{s}$ is introduced (moreover, this factor is
hardly different from 1). Hence,
the negative probability near the end-point is largely driven by the magnetic
moment term $\hat{\lambda}_2$.  
\item 
A comparison of the dilepton mass spectrum resulting from 
eq.~(\ref{eqn:dbds0}) of this work and eq.~(3.21) in FLS \cite{falketal}
(i.e. eq. (54) given above)
is shown in Fig.~\ref{fig:flvsus}, where we have used the 
input parameters given in Table \ref{parameters},
except that we have set $m_s=0$ to conform to the limit in which these two 
equations are derived. The two
curves differ in the large $\hat{s}$ region with ours becoming negative
before the kinematic end-point is actually reached. 
\end{itemize} 
 The  normalized FB asymmetry, $d\bar{\cal A}(\s)/d\s$,
in the HQE-approach and the parton model are shown in 
Fig.~\ref{fig:hillerfig2}. We find that this asymmetry is stable against 
leading order power corrections up to $\s \leq 0.6$, but the corrections
become increasingly large and eventually uncontrollable due to the 
unphysical
behaviour of the HQE-based dilepton mass distribution as $\s$ approaches 
$\s^{max}$ (see Fig.~\ref{fig:flvsus}).
Based on these investigations, we must conclude that the HQE-based approach
has a  restrictive kinematical domain for its validity. In particular,
it breaks down for the high dilepton invariant mass region in \bxsll.

\begin{table}[h]
	\begin{center}
	\begin{tabular}{|l|l|}
	\hline
	\multicolumn{1}{|c|}{Parameter}	& 
		\multicolumn{1}{|c|}{Value}	\\
	\hline \hline
	$m_W$			& $80.26$ (GeV)	\\
	$m_Z$			& $91.19$ (GeV)	\\
        $\sin^2 \theta_W $      & $0.2325$ \\
	$m_s$			& $0.2$ (GeV)	\\
	$m_c$			& $1.4$ (GeV) \\
	$m_b$			& $4.8$ (GeV) \\
	$m_t$			& $175 \pm 9$ (GeV)	\\
	$\mu$			& $5^{+5.0}_{-2.5}$ (GeV)	\\
	$\Lambda_{QCD}^{(5)}$	& $0.214^{+0.066}_{-0.054}$ (GeV)	\\
	$\alpha_{QED}^{-1}$	& 129		\\
	$\alpha_s (m_Z) $	& $0.117 \pm 0.005$ \\
	${\cal B}_{sl}$		& $(10.4 \pm 0.4)$ \% 	\\
        $\lambda_1$             & $-0.20$ (GeV$^2$) \\   
        $\lambda_2$             & $+0.12$ (GeV$^2$) \\
	\hline
	\end{tabular}
	\end{center}
\caption{Values of the input parameters used in the numerical
          calculations of decay rates. Unless, otherwise specified,
          we use the central values.} 
\label{parameters}
\end{table}

\begin{table}[h]
	\begin{center}
	\begin{tabular}{|l|l|}
	\hline
	\multicolumn{1}{|c|}{Coefficient}	& 
		\multicolumn{1}{|c|}{Value}	\\
	\hline \hline
	$C^{(0)}$			& $+0.3805$ 	\\
	$C_7^{\mbox{eff}}$		& $-0.3110$ 	\\
        $C_9^{NDR}$             & $+4.1530$ \\
	$C_{10}$			& $-4.5461$ 	\\
	\hline
	\end{tabular}
	\end{center}
\caption{ Wilson coefficients used in the numerical
          calculations corresponding to the central values 
          given in Table \protect\ref{parameters}.}
\label{wilson}
\end{table}

  This behaviour of the dilepton mass spectrum in \bxsll is not unexpected,
as similar behaviours have been derived near the end-point of the lepton
energy spectra in the decays $B \to X \ell \nu_\ell$
in the HQE approach \cite{manoharwise}.
 To stress these similarities, we show the 
power correction in the dilepton 
mass distribution as calculated in the HQE approach compared to the parton 
model through the ratio defined as:
\begin{equation}
 R^{\mbox{\small HQE}} (\s) \equiv \frac{{\rm d}{\cal 
B}/d\hat{s}(\mbox{HQE}) - 
{\rm d}{\cal B}/d\hat{s}(\mbox{Parton Model})}{ {\rm d}{\cal
B}/d\hat{s}(\mbox{Parton Model})}
\label{ratiohqepm}
\end{equation}

\begin{figure}[htb]
\vskip -2.0truein
\centerline{\epsfysize=7in
{\epsffile{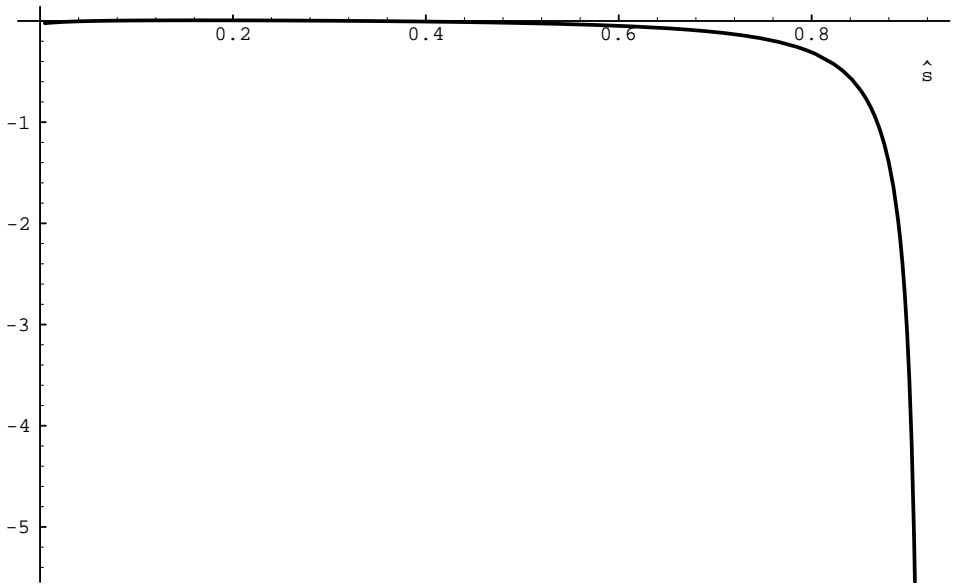}}}
\vskip -2.0truein
\caption[]{The correction factor $R^{HQE}(s)$ (in percentage) as defined in
eq. (\protect\ref{ratiohqepm}) for the dilepton mass spectrum ${\rm 
d}{\cal B} (B \to X_s \ell^+\ell^-)/{\rm d} \hat{s}$.
The parameters used are given in Table \ref{parameters}.}
\label{fig:hillerfig3}
\end{figure}

 The correction factor  $R^{\mbox{HQE}} (\s)$ for \bxsll
shown in Fig.~\ref{fig:hillerfig3} is qualitatively similar to 
the corresponding factor in the lepton
energy spectrum in the decay $ B \to X_c \ell \nu_\ell$, given in Fig. 6
of \cite{manoharwise}.
 Finally, we note that we have been
able to derive
the power corrected rate for the semileptonic decays $B \to X_c \ell
\nu_\ell$ obtained by Manohar and Wise in \cite{manoharwise},
taking the appropriate limits of our calculations and taking into
account the differences in our normalization of states and conventions,
as shown in Appendix C.

  Finally, since the HQE-improved expression for the decay rate including 
the $s$-quark mass effects is rather long, we give below the results
in a numerical form:
\begin{equation}
\Gamma^{\mbox{\small HQE}}=\Gamma^{b} (1 + C_1\hat{\lambda}_1 + C_2 
\hat{\lambda}_2 )\, ,
\label{gammahqet}
 \end{equation}
where $\Gamma^{b}$ is the parton model decay width for \bsll and the
coefficients depend on the input parameters. For the central values of
the parameters given in Table \ref{parameters}, they have the values
$C_1 =0.501$ and $C_2 = -7.425$. This leads to a reduction in the decay
width by $-4.1\%$, using the values of $\lambda_1$ and $\lambda_2$ given
in Table \ref{parameters}. Moreover, this reduction is mostly contributed by 
the $\lambda_2$-dependent term. We recall that the coefficient of the 
$\hat{\lambda}_1$ term above is the same as in the semileptonic width 
$\Gamma(B \to X_u \ell \nu_\ell)$,
but the coefficient of the $\hat{\lambda}_2$ term above is larger than
the corresponding coefficient $(=-9/2)$ in the semileptonic decay width.
Hence, the power corrections in $\Gamma(B \to X_u \ell \nu_\ell)$ and
$\Gamma(\mbox{\bxslll})$ are rather similar but not identical. 
\section{$B$-meson Wave function Effects 
in \bxsll} \label{wavefunction}
  In this section, we present our estimates of the non-perturbative
effects on the decay distributions in \bxslll. These effects are connected
with the bound state nature of the $B$ hadron and the
physical threshold in the \bxsll in the final state.
In order to implement these effects on the decay 
distributions in \bxslll, we resort to
the Gaussian Fermi motion model introduced in \cite{aliqcd}.  
In this model,
the $B$-meson consists of a $b$-quark and an spectator antiquark $\bar{q}$ and
the four-momenta of the constituents are required to add up to the
four-momentum of the $B$-meson.
In the rest frame of the $B$-meson the $b$-quark and the spectator move
back-to-back with three momenta $\vec{p}_b=-\vec{p}_q \equiv \vec{p}$.
Energy conservation then implies the equation
\[ m_B = \sqrt{m_b^2 + \vec{p}^2} + \sqrt{m_q^2 + \vec{p}^2} \quad, \]
which can only hold for all values of $|\vec{p}|$,
if at least one of the
masses becomes momentum dependent. We treat the spectator quark
$m_q$ as a 
 momentum-independent parameter; the $b$-quark mass is then
momentum dependent and we
 denote it by $W(p)$:
\begin{equation}
\label{lett14}
W^2(p) = {M_B}^2 + {m_q}^2 -2M_B \sqrt{p^2 + {m_q}^2} \quad .
\end{equation}
The $b$-quark, whose decays determine the dynamics, is given
 a non-zero momentum having a Gaussian
distribution, with the width determined by the
parameter $p_F$:
\begin{equation}
\label{lett13}
 \phi(p)= \frac {4}{\sqrt{\pi}{p_F}^3} \exp (\frac {-p^2}{{p_F}^2})
\quad ; \quad p = |\vec{p}| ~,
\end{equation}
with the normalization
$ \int_0^\infty \, dp \, p^2 \, \phi(p) = 1 $. 
 The distributions from
 the decay of the $B$-meson at rest are then obtained by
convoluting the appropriately boosted 
partonic distributions with the Gaussian distribution. The resulting 
spectra and decay rates
depend essentially on two parameters, $p_F$, determining the 
non-perturbative width of the momentum distribution, and $m_q$ (or 
equivalently $W(p)$), which determines the height.
 In the Fermi motion
model, the problem of negative probabilities encountered in the HQE approach
for the high dilepton masses near $s \to s_{max}$ is not present, which
motivates us to use
this model as a reasonable approximation of the non-perturbative effects  
in the entire dilepton mass range. The success of this model in describing
the inclusive lepton energy spectra in $B \to (X_c,X_u) \ell \nu_\ell$
and \bxsg strengthens this hope.

     In the decay \bxslll, the distribution
$d {\cal B}/d\hat{s}$ depends on the Lorentz-invariant variable $\hat{s}$ 
only. So, the Lorentz boost involved in the Fermi motion 
model (Doppler shift) leaves the dilepton mass distribution invariant. 
However, since the $b$-quark mass $W(p)$ is 
now a momentum-dependent quantity, this distribution is affected due to the
 difference ($W(p)-m_b)$ (mass defect), which 
rescales the variable $\hat{s}$ and hence smears the dilepton distribution
calculated in the parton model.
 For different choices of the model
parameters $(p_F,m_q)$ corresponding to the same effective $b$-quark mass,
$\langle W \rangle$, the dilepton mass distributions should be very similar
\cite{Greub96}, 
which indeed is the case as we have checked numerically but do not show
the resulting distributions here.

  The situation with the FB asymmetry (or the energy asymmetry) is, however,
quite different. Being an angular-dependent quantity,  it is not
Lorentz-invariant and is sensitive to both the Doppler shift and the 
mass defect. We give in Appendix E, the Dalitz distribution $d^2\Gamma(B \to
X_s \ell^+ \ell^-)/ds du$ in the Fermi motion model,  given the partonic 
double distribution $d^2 \Gamma (\mbox{\bsll}) /d\hat{s} d \hat{u}$
in the $b$-quark rest frame. These details, hopefully, will be useful 
in the
analysis of data in \bxsll due to the popularity of the Fermi motion model.

As we calculate the
branching ratio for the inclusive decay \bxsll in terms of the
semileptonic decay branching ratio ${\cal B} (B \to X\ell \nu_\ell)$, we
have to correct the normalization due to the variable $b$-quark mass in
both the decay rates. To get the decay rates in this model
one  first implements the wave function effects and then  integrates the 
spectra.
Fixing $m_b$ but varying the model parameters $p_F$ and $m_q$ yields 
variable effective (momentum-dependent) $b$-quark mass $\langle W \rangle$.
We recall that the decay widths for \bxsll and $B \to X
\ell \nu_\ell$ in this model are proportional to $\langle W^5 \rangle$
\cite{ag1}. Hence the  decay widths for both the decays
individually are rather sensitive
to  $\langle W \rangle$. This dependence largely (but not exactly) cancels
out in the branching ratio ${\cal B}(\mbox{\bxslll})$. Thus, varying
$\langle W \rangle$ in the range  $\langle W \rangle= 4.8 \pm 0.1$ GeV
results in $\Delta \Gamma(\mbox{\bxslll})/\Gamma = \pm 10.8\%$.
 However, the change in the
branching ratio itself is rather modest, namely
$\Delta {\cal B}(\mbox{\bxslll})/{\cal B} = \pm 2.3\%$.
This is rather similar to what we have obtained in the HQE approach.

    The theoretical uncertainties in the branching ratios
for \bxsll from the 
perturbative part, such as the ones from the indeterminacy in the top 
quark mass,   
the QCD scale $\Lambda_{QCD}$ and the renormalization scale $\mu$,  
have been investigated in the literature \cite{Misiak1,buras}.
 We have
recalculated them for the indicated ranges of the parameters in
Table \ref{parameters}. The resulting (SD) branching ratios and their 
present uncertainties are found to be:
\begin{eqnarray}
{\cal B}(B \to X_s e^+ e^-) = (8.4 \pm 2.3) \times 10^{-6} \, , \nonumber\\ 
{\cal B}(B \to X_s \mu^+ \mu^-) = (5.7 \pm 1.2) \times 10^{-6} \, , 
\nonumber\\ 
{\cal B}(B \to X_s \tau^+ \tau^-) = (2.6 \pm 0.5) \times 10^{-7} \, , 
\end{eqnarray}
where in calculating the branching ratio ${\cal B}(B \to X_s \tau^+ \tau^-)$,
we have included the $\tau$-lepton mass terms in the matrix element
\cite{KS96}.
These uncertainties, typically $\pm 25\%$, are
much larger than the wave-function-dependent uncertainties, and so the
theoretical accuracy of the SD-part
in the SM in these decays is not compromised by the non-perturbative 
effects.
 
\begin{figure}[htb]
\vskip -0.1truein
\centerline{\psfig{figure=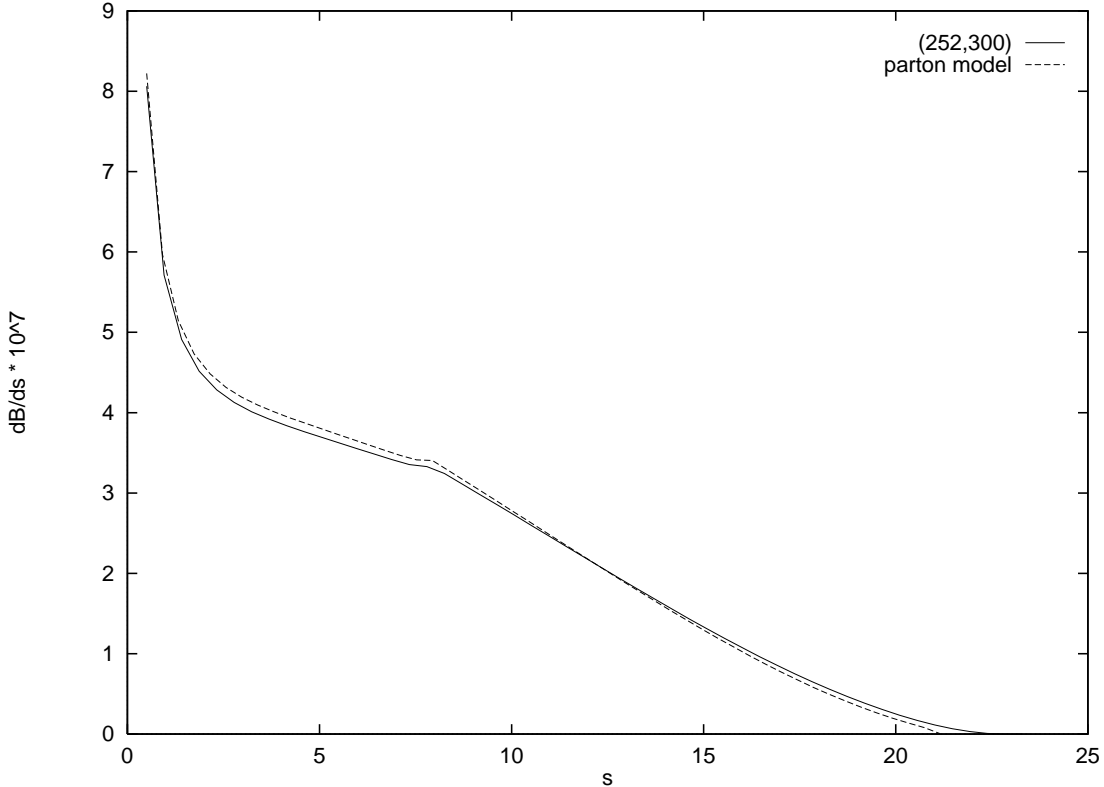,height=15.0cm,angle=270}}
\vskip -0.1truein
\caption[]{Differential branching ratio $d {\cal B}/ds$ for
$B \to X_s \ell^+ \ell^-$ 
in the SM including the next-to-leading order QCD corrections.
 The dashed curve
corresponds to the parton model with the parameters given in Table 1 and the
solid curve results from the Fermi motion model with the model parameters
$(p_F,m_q)=(252,300)$ MeV, yielding an effective $b$-quark mass
 $\langle W \rangle =4.85$ GeV.}
 \label{fig:fermi1}
\end{figure}

\begin{figure}[htb]
\vskip -0.5truein
\centerline{\psfig{figure=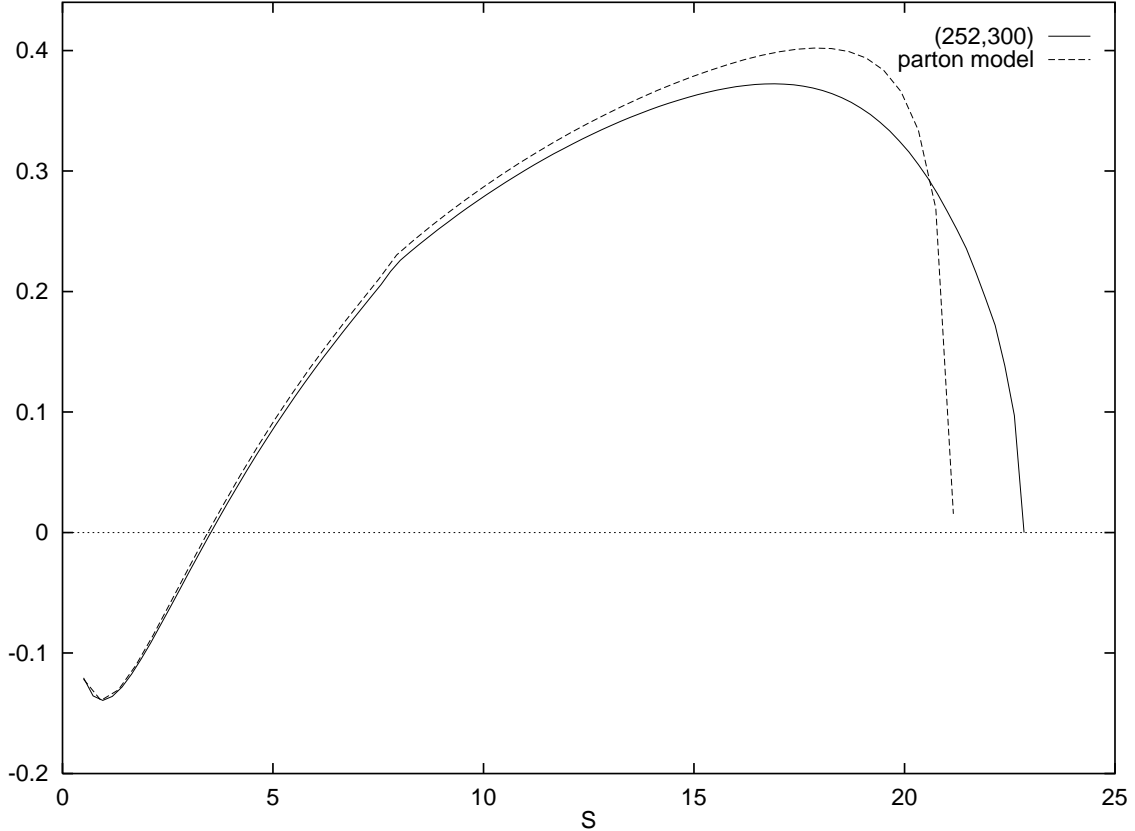,height=16.0cm,angle=270}}
\vskip -0.1truein
\caption[]{Normalized differential FB asymmetry $d\overline{{\cal A}}(s)/ds$
for $B \to X_s \ell^+ \ell^- $ in the SM including the
next-to-leading order QCD correction. The dashed curve corresponds to the 
parton model and the solid curve results from the Fermi motion model 
with the model parameters $(p_F,m_q)=(252,300)$ MeV, yielding an 
effective $b$-quark mass $\langle W \rangle =4.85$ GeV.}
\label{fig:fermi2}
\end{figure}

 We show the resulting dilepton invariant mass 
distribution in 
Fig.~\ref{fig:fermi1} and the FB-asymmetry in Fig.~\ref{fig:fermi2}, where
for the sake of illustration we have
used  the values $(p_F, m_q)=(252,300)$ (both in MeV), which  
correspond to an allowed set of 
parameters obtained from the analysis of the measured photon energy spectrum
in \bxsg, using the same model \cite{ag2}. We see that the dilepton mass 
distribution is very stable against Fermi motion effects over most part
of this spectrum, as expected. The end-point spectrum in this 
model extends to the physical kinematic limit in \bxslll,
$ s^{max}= (m_B-m_K)^2$, which obtains for $m(X_s) = m_K$, as opposed to the
parton model, in which $s^{max}=(m_b-m_s)^2$.  
The two thresholds can be made
to coincide for only unrealistically high values of $m_b$ and $m_s$.
The FB-asymmetry shows a more marked dependence on the model parameters,
which becomes very significant in the high dilepton mass region.

As the parameters of the Fermi motion model are not presently very 
well-determined from the fits of the existing data \cite{ag2,CLEOslfm},
one has to vary these parameters and estimate the resulting dispersion on
the distributions in \bxslll. We show in Figs.~7 and 8 the dilepton mass
distribution and the FB asymmetry, respectively, indicating also the ranges
of the parameters $(p_F,m_q)$. The resulting theoretical 
uncertainty in the distributions is found to be modest.

\begin{figure}[htb]
\vskip -0.5truein
\centerline{\psfig{figure=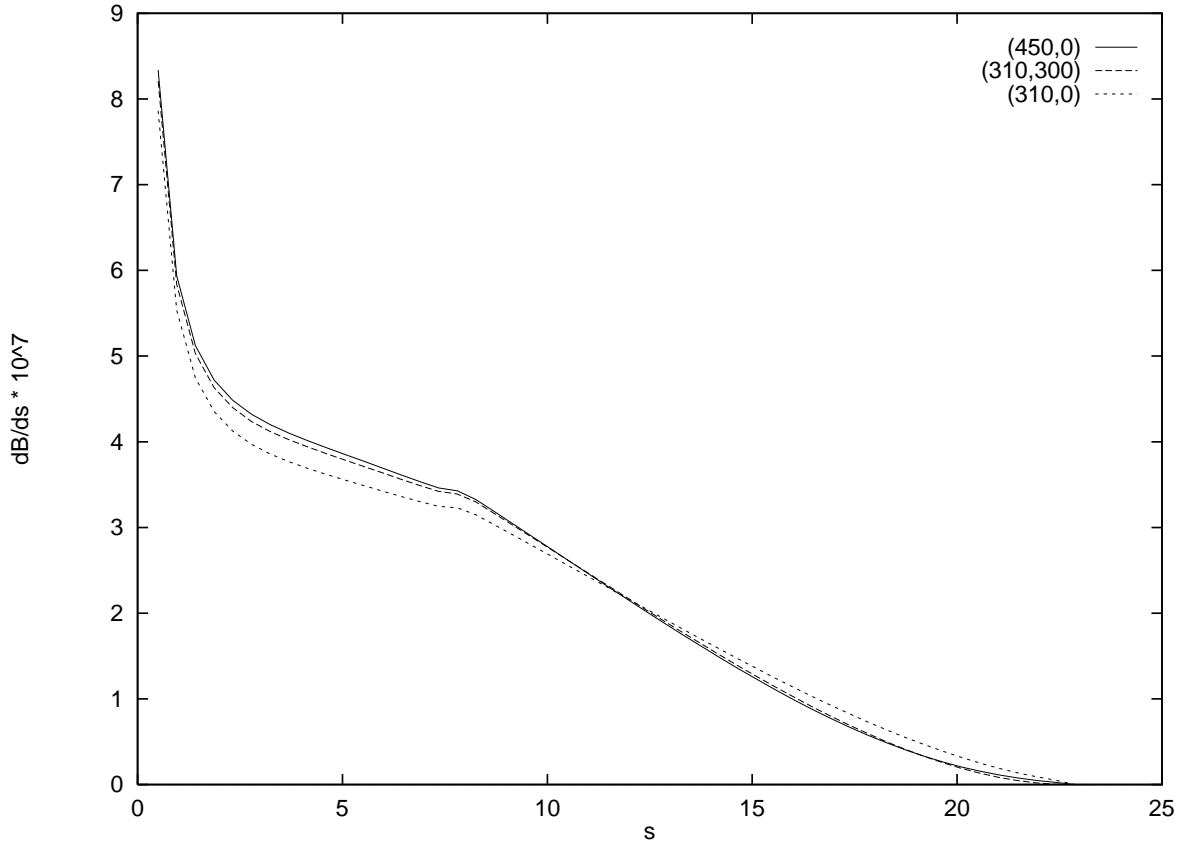,height=16.0cm,angle=270}}
\vskip -0.1truein
\caption[]{Differential branching ratio $d {\cal B}/ds$ for
$B \to X_s \ell^+ \ell^-$
using the Fermi motion model for three 
different pairs of the 
model parameters $(p_F,m_q)=(450,0)$ MeV (solid curve), $(310,300)$ MeV
(long dashed curve), and $(p_F,m_q)=(310,0)$ MeV (short dashed curve)
 yielding the effective $b$-quark masses
 $\langle W \rangle =4.76$ GeV, $4.80$ GeV, and $4.92$ GeV, respectively.}
\label{fig:fermi5}
\end{figure}
\begin{figure}[htb]
\vskip -0.5truein
\centerline{\psfig{figure=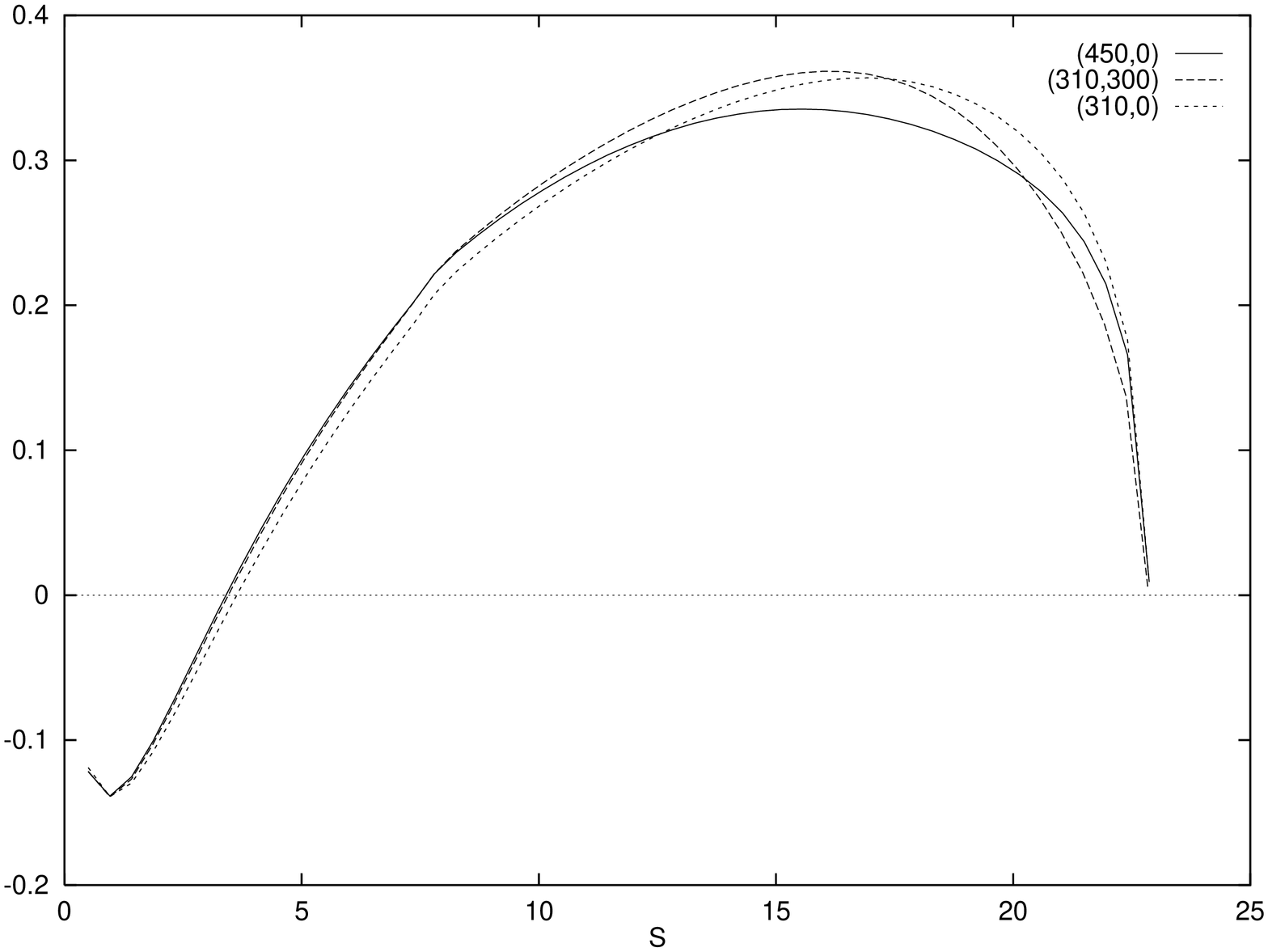,height=16.0cm,angle=270}}
\vskip -0.1truein
\caption[]{Normalized differential FB asymmetry $d\overline{{\cal A}}(s)/ds$
for $B \to X_s \ell^+ \ell^- $
using the Fermi motion model for three
different pairs of the
model parameters $(p_F,m_q)=(450,0)$ MeV (solid curve), $(310,300)$ MeV
(long dashed curve), and $(p_F,m_q)=(310,0)$ MeV (short dashed curve)  
 yielding the effective $b$-quark masses
 $\langle W \rangle =4.76$ GeV, $4.80$ GeV, and $4.92$ GeV, respectively.}
\label{fig:fermi6}
\end{figure}
\begin{figure}[htb]
\vskip -0.5truein
\centerline{\psfig{figure=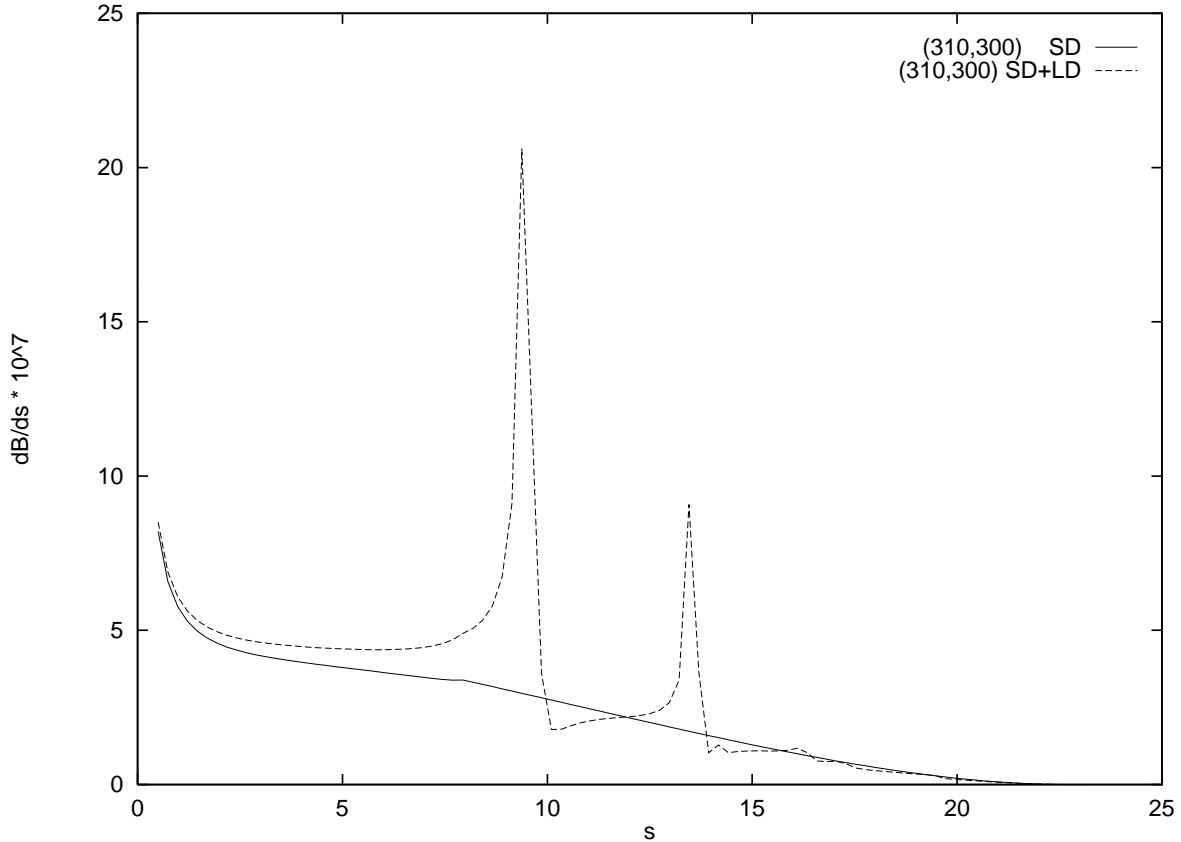,height=16.0cm,angle=270}}
\vskip -0.1truein
\caption[]{Differential branching ratio $d{\cal B}/ds$ for $B \to X_s \ell^+ 
\ell^- $ calculated in the SM using the
next-to-leading order QCD corrections and Fermi motion effect (solid 
curve), and including the LD-contributions (dashed curve).
The Fermi motion model parameters $(P_F,m_q)$ are displayed in the figure.
Note that the height of the $J/\psi$ peak is suppressed due to the linear
scale.} 
\label{fig:lddb}
\end{figure}
\begin{figure}[htb]
\vskip -0.5truein
\centerline{\psfig{figure=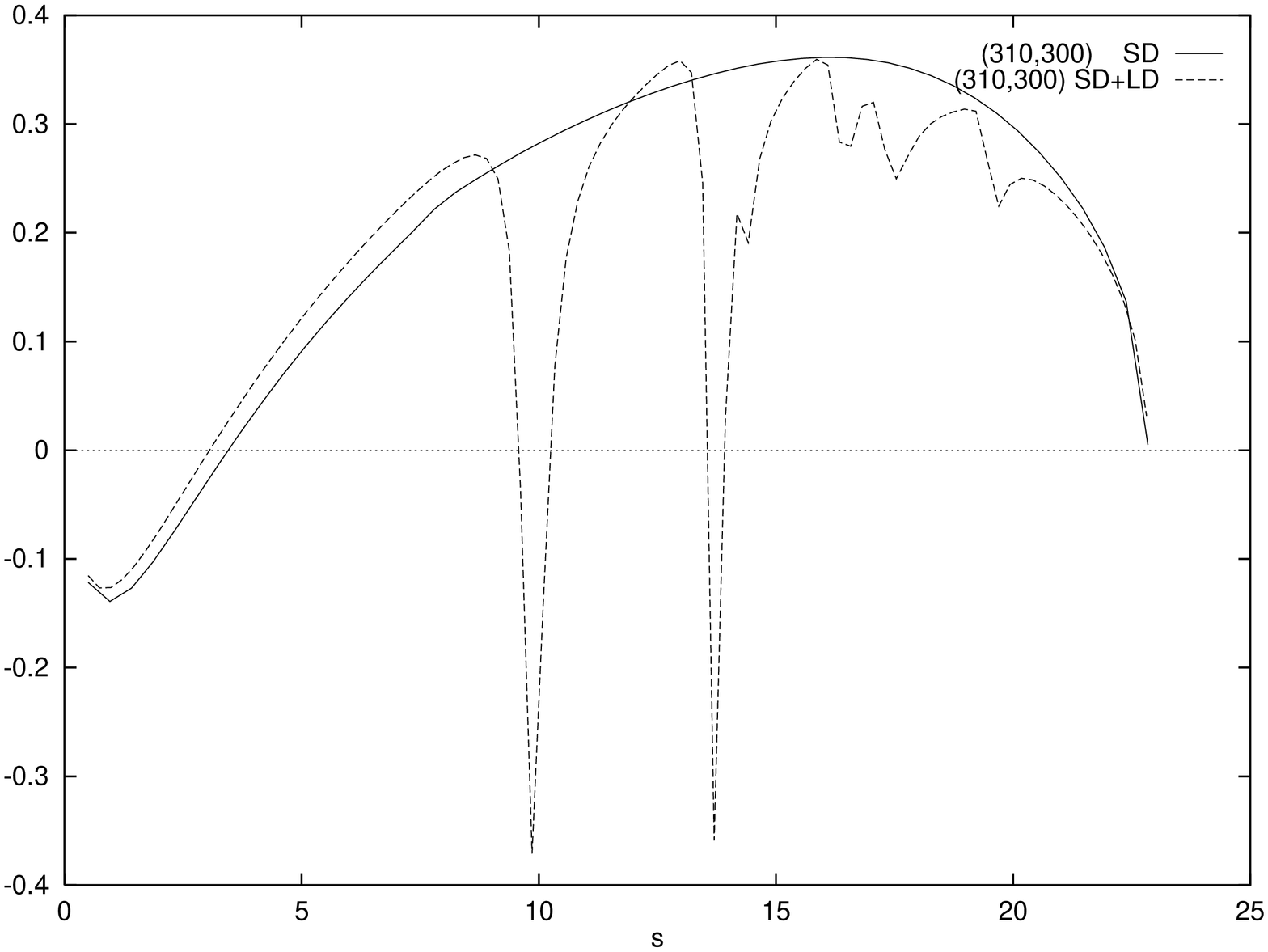,height=16.0cm,angle=270}}
\vskip -0.1truein
\caption[]{Normalized FB asymmetry $\overline{{\cal A}}(s)$
 for $B \to X_s \ell^+ \ell^- $ calculated in the SM 
using the next-to-leading order QCD corrections and Fermi motion 
effect (solid
curve), and including the LD-contributions (dashed curve).
The Fermi motion model parameters $(P_F,m_q)$ are displayed 
in the figure.}
\label{fig:ldda}
\end{figure}
\section{\bf LD contributions in \bxsll}

Next, we implement the effects of LD contributions in the
processes \bxslll. The issues involved here have been discussed recently in 
\cite{LW96}-\cite{Ahmady96} and so we will be short in this part.
The LD contributions due to the vector 
mesons $J/\psi$ and $\psi^\prime$ and higher resonances, as well as the
$(c\bar{c})$ continuum contribution, which we have already included in
the coefficient $C_9^{\mbox{eff}}$, 
appear in the $(\bar{s}_L \gamma_\mu b_L)(\bar{e} \gamma^\mu e)$
interaction term only, i.e. in the coefficient of the operator $O_9$.
 This implies that such LD-contributions should change
$C_9$ effectively, but keep $C_7^{\mbox{eff}}$ and $C_{10}$ unchanged. In principle,
one has also a LD contribution in the effective coefficient $C_7^{\mbox{eff}}$;
this, however, has been discussed extensively in the context of the \bxsg 
decay and  estimated to be 
small \cite{bsgamld,Deshpande96}. The LD-contribution is negligible in 
$C_{10}$. Hence, the three-coefficient fit of the data on \bxsll 
and \bxsg, proposed in \cite{agm94} on the basis of the SD-contributions,
can be carried out also including the LD-effects.
In accordance with this, to incorporate the LD-effects in \bxslll, the 
function $Y(\s)$ introduced earlier is replaced by, \begin{equation}
	Y(\s) \rightarrow Y^\prime(\s) \equiv Y(\s) + 
		Y_{res}(\s),
\end{equation}
where $Y_{res}(\s)$ is given as \cite{amm91},
\begin{equation}
	Y_{res}(\s) = \frac{3}{\alpha^2} \kappa \, C^{(0)}
		\sum_{V_i = \psi(1s),..., \psi(6s)}
		\frac{\pi \, \Gamma(V_i \rightarrow l^+ l^-)\, M_{V_i}}{
		{M_{V_i}}^2 - \s \, {m_b}^2 - i M_{V_i} \Gamma_{V_i}} ,
\label{LDeq}
\end{equation}
and
\begin{equation}
	C^{(0)} \equiv 3 C_1^{(0)} + C_2^{(0)} + 3 C_3^{(0)} + 
		C_4^{(0)} + 3 \, C_5^{(0)} + C_6^{(0)} .
\end{equation}
Here we adopt $\kappa = 2.3 $
for the numerical calculations \cite{LW96}.
Of course, the data determines only the combination 
$\kappa \, C^{(0)} = 0.88$. The relevant parameters of the charmonium 
resonances $(^1S,...,^6S)$ are given in the Particle Data Group
\cite{PDG96}, and we have averaged the leptonic widths for the decay modes
$V \to \ell^+ \ell^-$ for $\ell=e$ and $\ell=\mu$. Note that in
extrapolating the dilepton masses away from the
resonance region, no extra $q^2$-dependence
is included in the $\gamma^{*}(q^2)$-$V_i$ junction.
(The $q^2$-dependence written explicitly in eq.~(\ref{LDeq})
is due to the Breit-Wigner shape of the resonances.)
This is an assumption and it 
may lead to an underestimate of the LD-effects in the low-$s$
region. However, as the present phenomenology is not equivocal on this
issue, any other choice at this stage would have been on a similar footing.
The resulting dilepton mass spectrum and
the FB asymmetry are shown in Fig.~\ref{fig:lddb} and
Fig.~\ref{fig:ldda}, respectively. The two curves labeled SD and SD$+$LD 
 include:
\begin{itemize}
\item Explicit $O(\alpha_s)$-improvement, calculated in the parton model
\cite{Misiak1,buras}.
\item Non-perturbative effects related with the bound state nature of the
$B$ hadrons and the physical threshold in the final state in \bxslll,
using the Fermi motion model \cite{aliqcd} with the parameters specified
in the figures.
\end{itemize}
In addition, the SD$+$LD case also includes the LD-effects due to the 
vector
resonances, contributing to $C_9^{\mbox{eff}}$ as discussed earlier.
 
\begin{figure}[htb]
\vskip -0.5truein
\centerline{\psfig{figure=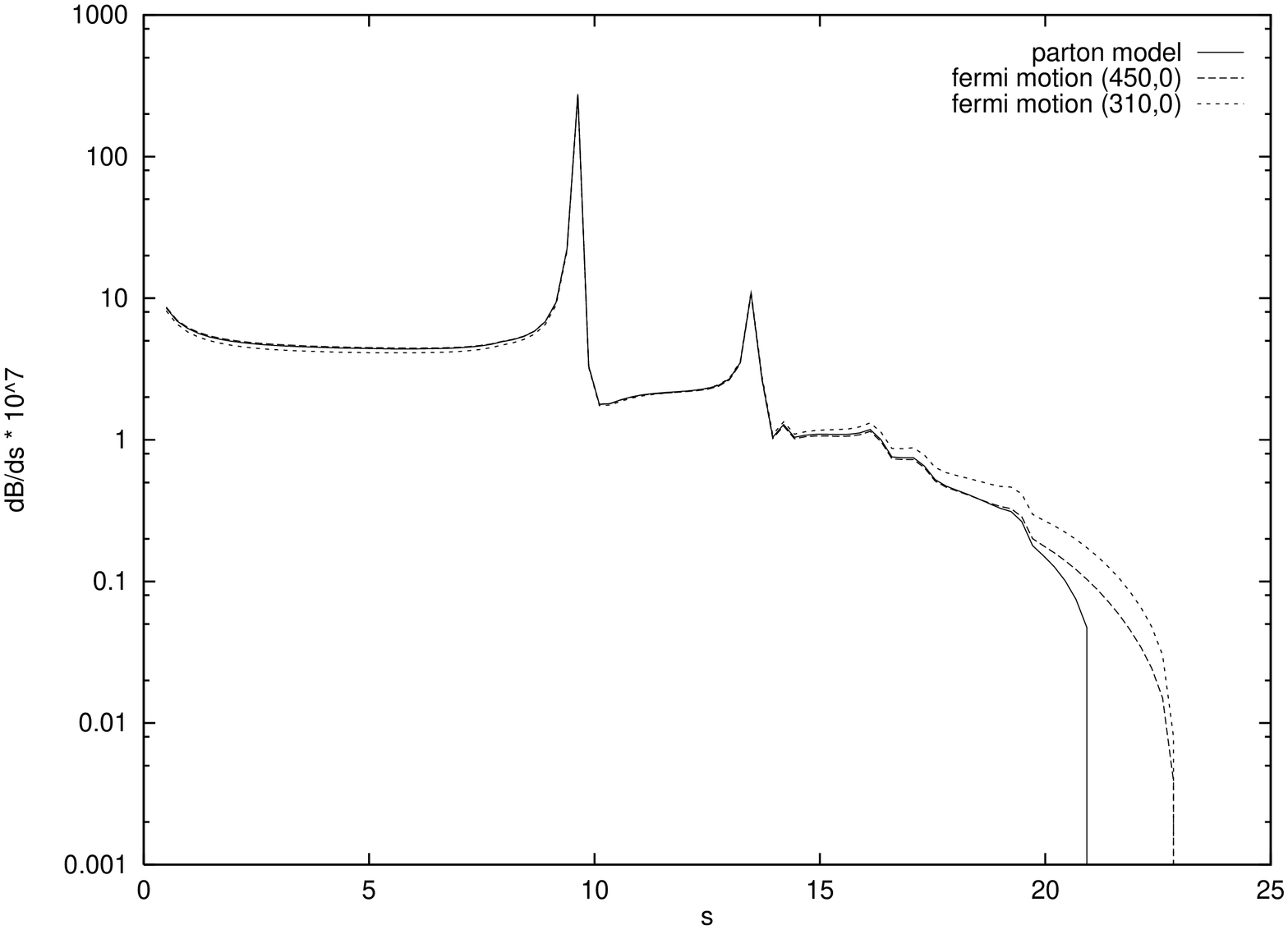,height=16.0cm,angle=270}}
\vskip -0.1truein
\caption[]{
Dilepton invariant mass distribution in $B \to X_s \ell^+ \ell^-$ in 
 the SM including
next-to-leading order QCD correction and LD effects. The solid curve
corresponds to the parton model and the short-dashed and long-dashed
curves correspond to including the Fermi motion effects. The
values of the Fermi motion model are indicated in the figure.}
\label{fig:dbrnsm}
\end{figure}

\begin{figure}[htb]
\vskip -0.5truein 
\centerline{\psfig{figure=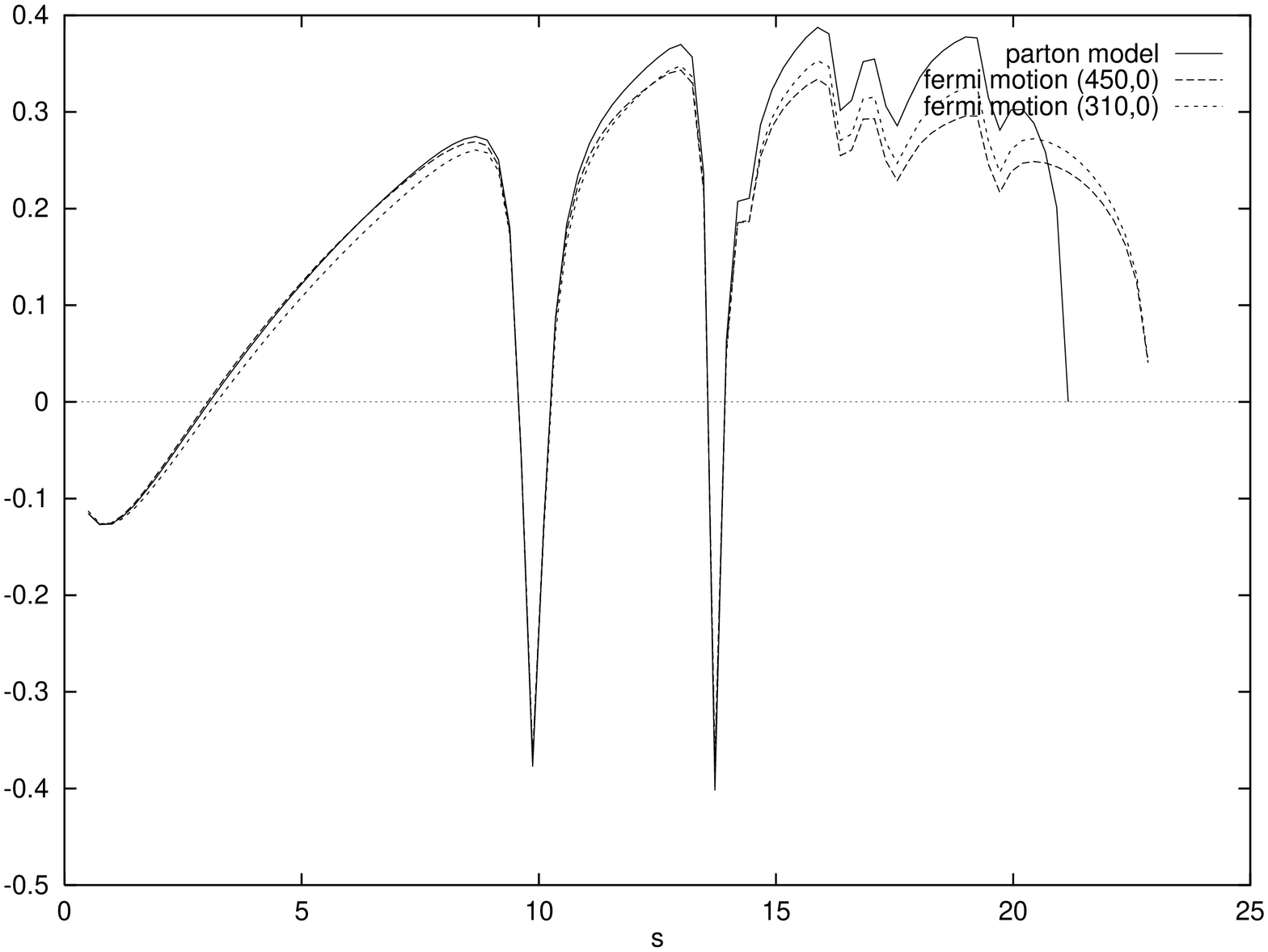,height=16.0cm,angle=270}}
\vskip -0.1truein 
\caption[]{FB asymmetry for $B \to X_s \ell^+ \ell^-$ in the SM
as a function of the dimuon invariant mass  including the
                next-to-leading order QCD correction and LD effects.
 The solid curve
corresponds to the parton model and the short-dashed and long-dashed
curves correspond to including the Fermi motion effects. The
values of the Fermi motion model are indicated in the figure.} 
\label{fig:asymmnsm}
\end{figure}

 Finally, the parametric dependence due to the Fermi motion model 
is shown in Figs.~\ref{fig:dbrnsm} and \ref{fig:asymmnsm} for the dilepton
mass spectrum and the FB asymmetry, respectively, and compared with the
case of the parton model in which case no wave-function effects are 
included. 
 These figures give a fair 
estimate of the kind of uncertainties present in these distributions from 
non-perturbative
effects. In particular, we draw attention to the marked dependence of the FB 
asymmetry to both the LD-(resonances) and wave function effects,
which is particularly noticeable in the region $m_{\ell \ell} > m(\psi 
^\prime)$. The dilepton invariant mass spectrum, on the other hand, is 
very stable except at the very end of the spectrum, which is clearly
different in all three cases shown.
%
%
\section{Concluding Remarks}
We have investigated the question of power corrections to the decay rates
and distributions in the FCNC process $B \to X_s \ell^+ \ell^-$ in the
HQE framework. Our motivation here was to check if indeed the entire
dilepton mass spectrum in these decays is calculable in a theoretically
controlled sense, which the existing results suggested \cite{falketal}.
Our calculations of this distribution and the integrated rate do not agree 
with the ones obtained in FLS \cite{falketal}.
We have presented the details of our computations here, 
including the power corrections to the FB asymmetry not calculated 
earlier. In line with
the analogous calculations for the lepton and photon energy spectra in
radiative and semileptonic $B$ decays, we have found that the HQE approach
has a limited region of applicability
in describing the dilepton mass spectrum in \bxslll.
In the latter case, 
the use of the leading-order HQE approach in
the high-$s$ region results in unphysical distribution and hence can not
be used for comparison with data. Excluding the high-$s$ 
region, the power corrections to the dilepton mass spectrum and the
FB asymmetry are, however, found to be small.
The inclusive decay rate 
$\Gamma(b \to s \ell^+ \ell^-)$ receives small power correction
in the HQE approach, typically
($-4\%)$, which is similar to the one in
the semileptonic decay width $\Gamma(b \to u \ell \nu_\ell)$ but not
identical.
 
  Despite progress in some sectors, 
the problem of incorporating non-perturbative effects in weak decays
remains theoretically an intractable problem. In the present context, the
structure functions $T_i$'s entering the decay distributions in \bxsll
are not known from first principles in QCD. We have modeled the
non-perturbative effects in \bxsll 
using a popular Fermi motion model \cite{aliqcd},
which allows to incorporate $B$-hadron wave function
effects and the correct threshold in the final states. Since this  
model gives a good description of the existing data on the lepton and 
photon energy spectra in $B$ decays \cite{ag2,CLEOslfm}, we hope that
it describes similar non-perturbative effects in the decay \bxsll as well.
We have estimated the
dispersion on the theoretical predictions for the dilepton
mass and the FB asymmetry resulting from the present uncertainty in the model
parameters. This dispersion is marked in the high 
dilepton mass region in the FB asymmetry, but the dilepton mass spectrum is
remarkably stable. Hence,
 in the very high-$s$ region, non-perturbative effects are
important and have to be included in order to have a reliable comparison
of the SM-based distributions and data, as and when they become available.

  Finally, we have incorporated the LD-effects using data in $B$ decays and
the measured properties of the resonances in the charmonium sector. As
discussed in the literature, this is not sufficient to uniquely determine the
dilepton distributions away from these resonances.
 In that context, we note that the vector-meson-dominance (VMD)
approximation
of the old vintage \cite{Terasaki81} is often invoked to model the    
$q^2$-dependence of the $\gamma$-$V$ junction, $g_V(q^2)$.
This VMD-framework has been used to estimate
the LD-effects in \bxsg \cite{Deshpande96}.
Theoretical uncertainties from these aspects in the
dilepton mass distributions in \bxsll have been discussed in
\cite{Ahmady96}. We hope that data from HERA on the photoproduction of
$J/\psi$ and $\psi^\prime$ (and other resonances) can be used to implement
the $q^2$-dependence
of the effective vertices to eliminate (or at least reduce) the present 
theoretical uncertainty from this source. However, as the HERA-data on the
relevant processes       
$\gamma^{*} (q^2) + p \to J/\psi (\psi^\prime) + p$ are still preliminary
and a $q^2$-dependence
has not yet quantitatively been extracted \cite{ZEUSwarsaw96},
we do not attempt to undertake an improved treatment of the LD-effects
in \bxsll here. In view of
this, and the remaining theoretical uncertainties on the perturbative
part discussed in section 3, the distributions
shown in Figs.~(\ref{fig:dbrnsm}) and (\ref{fig:asymmnsm}) have an
overall uncertainty of order $\pm 25\%$.

{\Large \bf Acknowledgements}

We thank Christoph Greub for helpful discussions and correspondence on
implementing the Fermi motion effects.
L.T.H. and T.M. would like to thank  C.S. Kim, T. Onogi, M. Tanaka and  
A. Yamada for useful discussions.
L.T.H. would also like to thank the theory group at KEK for the
warm hospitality during his stay at KEK.
The work of L.T.H. is supported by a grant from the Ministry of
Education, Science and Culture (Mombusho), Japan. The research work of 
T.M. has been partially supported by the Deutscher Akademischer Austausch 
Dienst (DAAD) and the Japan Society for the Promotion of Science (JSPS),
which made a visit to DESY-Hamburg possible, where this work was completed.
%
%
%
\newpage
\centerline{\bf\underline{Appendices}\\}
\begin{appendix}
\renewcommand{\theequation}{\Alph{section}-\arabic{equation}}

\setcounter{equation}{0}
\section{The functions $T_{i}^{(j)}(v.\hat{q}, \hat{s})$}

In this appendix we list the expressions for
the decomposition of the structure functions $T_{i}(v.\hat{q},\hat{s}), 
(i=1,2,3)$ in terms of the functions 
$T_{i}^{(j)}(v.\hat{q}, \hat{s})$, with $j=0,1,2, s, g, \delta$, 
representing the power
corrections in \bsll up to and including terms of order $M_B/m_b^3$ and 
explicitly keeping the $s$-quark mass dependence. The 
origin of these individual terms is explained in the text. The parton model
contributions $T_{i}^{(0)}$ are  given in eqs.~(A-1) - (A-3). 

\begin{eqnarray}
        {{T_1}^{(0)}}^{L/R} & = & - \frac{1}{x} \frac{M_B}{m_b} \left\{
		(1 - \z ) | \cnt|^2 
		\right.
		\nonumber \\
	& & \left. 
		\; \; \; \; \; \; \; \; \; \; 
		+ \frac{4}{\s^2} \left[ 
		\left( 1 + \ms^2 \right)
			\left( 2 \, (\z)^2 - \s (\z) - \s \right) 
			- 2 \ms^2 \, \s \right] | C_7^{\mbox{eff}} |^2
		\right.
		\nonumber \\
	& & \left. 
		\; \; \; \; \; \; \; \; \; \; 
		+ \frac{4}{\s} \left[ 
		\z - \s - \ms^2 \, (\z) 
		\right] {\rm Re} \left[ ( \cnt )^\ast C_7^{\mbox{eff}} \right]
		\right\}
		\, , \\
        {{T_2}^{(0)}}^{L/R} & = & - \frac{2}{x} \frac{M_B}{m_b} \left\{
		|\cnt |^2
		- \frac{4}{\s} \left( 1 + \ms^2 \right) 
			|C_7^{\mbox{eff}} |^2
		\right\}
		\, , \\
        {{T_3}^{(0)}}^{L/R} & = & \frac{1}{x} \frac{M_B}{m_b} \left\{
		- |\cnt |^2 
		- \frac{4}{\s^2} \left[ 2 \, (\z) - \s\right]
			\left[ 1 - \ms^2 \right] |C_7^{\mbox{eff}}|^2
		\right.
		\nonumber \\
	& & \left. 
		\; \; \; \; \; \; \; \; \; \; 
		- \frac{4}{\s} \left( \ms^2 + 1	\right) 
		{\rm Re} \left[ (\cnt )^\ast C_7^{\mbox{eff}} \right] 
		\right\}
		\, , \\
        {{T_1}^{(1)}}^{L/R} & = & - \frac{1}{3} \mb \, 
		( \loo + 3 \lto ) \left\{
		\left[ \frac{1}{x} - \frac{2}{x^2} \left( \s - (\z)^2 
			\right) \right] | \cnt |^2
		\right. \\
	& & 
		+ \frac{4}{\s^2} \left[ 
		\frac{1}{x} \left( \s - 2 \, (\z)^2 \right) 
		- \frac{2}{x^2} \left( \s^2 - 2 \, \s (\z) - \s (\z)^2 
			+ 2 \, (\z)^3 \right) \right]\; (1+\ms^2) | C_7^{\mbox{eff}} |^2
\nonumber \\
& &   \left.
  - (\s-\z^2)\; \ms^2 \frac{8}{\s x^2}\; {\rm Re} (\cnt )^\ast C_7^{\mbox{eff}}
                \right\} \, , 
\nonumber \\
        {{T_2}^{(1)}}^{L/R} & = & - \frac{2}{3} \mb \, 
		( \loo + 3 \lto ) \left[
		\frac{1}{x} + \frac{2}{x^2} \, \z \right]
		\left[ - | \cnt |^2 + \frac{4}{\s} \; (1+\ms^2) | C_7^{\mbox{eff}} |^2
                \right] \, , \\
        {{T_3}^{(1)}}^{L/R} & = & - \frac{2}{3} \mb \, 
		( \loo + 3 \lto ) \left\{
		\frac{1}{x^2} ( 1 - \z ) | \cnt |^2    
		\right.
		\nonumber \\
	& &  
		- \frac{4}{\s^2} \left[ \frac{1}{x} \z 
		- \frac{1}{x^2} \left( \s + \s (\z) - 2 \, (\z)^2 \right)
		\right] \; (1-\ms^2) | C_7^{\mbox{eff}} |^2
\nonumber \\
& & \left. 
- \z \; \ms^2 \frac{4}{\s x^2}\; {\rm Re} (\cnt )^\ast C_7^{\mbox{eff}}
                \right\} \, , \\
        {{T_1}^{(2)}}^{L/R} & = & \frac{1}{3} \mb \, \loo 
		\left\{ 
		\left[ - \frac{4}{x^3} \left( \s - (\z)^2 \right)
			+ \frac{3}{x^2} \right] (1 - \z) 
			| \cnt |^2 
		\right.
		\nonumber \\
	& & - \frac{4}{\s^2 x^3} \left[
-4 \s^2 -12 \ms^2 \s^2 + 3 \s x + 9 \ms^2 \s x -4 \s^2 \z -4 \ms^2 \s^2 \z +7 \s x \z
\right.
\nonumber \\
& &
 + 7 \ms^2 \s x \z + 12 \s \z^2+ 20 \ms^2 \s \z^2 -6 x \z^2 -6 \ms^2 x \z^2 + 4 \s \z^3
\nonumber \\
& & \left. + 4 \ms^2 \z^3 \s -4 x \z^3 -4 \ms^2 x \z^3 -8 \z^4 -8 \ms^2 \z^4
		\right] 
		 | C_7^{\mbox{eff}} |^2
		\nonumber \\
	& & 
		+ \frac{4}{\s x^3} \left[
4 \s^2 - 5 \s x - 4 \s \z +4 \ms^2 \s \z +3 x  \z-3 \ms^2 x \z -4 \s \z^2+2 x \z^2 
\right.
\nonumber \\
& & \left. \left. +4 \z^3 -4 \ms^2 \z^3
		\right]
		{\rm Re} \left[ ( \cnt )^\ast \, C_7^{\mbox{eff}} \right]
                \right\} \, , \\
        {{T_2}^{(2)}}^{L/R} & = & - \frac{2}{3} \mb \, \loo 
		\left[ \frac{4}{x^3} \left( \s - (\z)^2 \right)
			- \frac{3}{x^2} - \frac{2}{x^2} \, \z \right]
                \nonumber \\
        & &    \left( | \cnt |^2 - \frac{4}{\s} (1+\ms^2)\; |C_7^{\mbox{eff}} |^2
		\right) \, , \\
        {{T_3}^{(2)}}^{L/R} & = & - \frac{1}{3} \mb \, \loo 
		\left\{
		\left[ \frac{4}{x^3} \left( \s - (\z)^2 \right)
			- \frac{5}{x^2} \right] | \cnt |^2 
		\right.
		\nonumber \\
	& & 
		+ \frac{4}{\s^2 x^3} \left[
-4 \s^2 +5 \s x + 8 \s \z  -6 x \z + 4 \s \z^2-4 x  \z^2 -8 \z^3
		 \right]  (1-\ms^2) \, | C_7^{\mbox{eff}} |^2
		\nonumber \\
	& & \left. 
		+ \frac{4}{\s x^3} \left[
(4 \s -3 x -4 \z^2)(1+\ms^2) -2 x \z 
		 \right]
		{\rm Re} \left[ ( \cnt )^\ast \, C_7^{\mbox{eff}} \right]
		\right\} \, , \\
        {{T_1}^{(s)}}^{L/R} & = & \frac{2}{\s \, x}
		\mb \, ( \loo + 3 \lto ) \left[ ( \s - \z ) 
		{\rm Re} \left[ ( \cnt )^\ast \, C_7^{\mbox{eff}} \right] +2 \ms^2 |C_7^{\mbox{eff}}|^2 \right] \, , \\
        {{T_2}^{(s)}}^{L/R} & = & 0 \, , \\
        {{T_3}^{(s)}}^{L/R} & = & - \frac{2}{\s \, x}
		\mb \, ( \loo + 3 \lto ) 
		{\rm Re} \left[ ( \cnt )^\ast \, C_7^{\mbox{eff}} \right] \, , \\
	{{T_1}^{(g)}}^{L/R} & = & \frac{1}{x^2} \mb \, \lto 
		\left\{ -( 1 - \z ) | \cnt |^2 
\right.
\nonumber \\
& & 
		+ \frac{4}{\s^2} \left[ \s +3 \ms^2 \s + \s (\z) (1+\ms^2) - 2 \, (\z)^2 \; (1+\ms^2) \right] | C_7^{\mbox{eff}} |^2 
		\nonumber \\
	& & \left. 
		+ \frac{4}{\s} ( \s - \z \; (1-\ms^2) ) 
		{\rm Re} \left[ ( \cnt )^\ast \, C_7^{\mbox{eff}} \right]
		\right\}
		\, , \\
        {{T_2}^{(g)}}^{L/R} & = & \frac{-2}{x^2} \mb \, \lto
                \left\{ - | \cnt |^2 - \frac{4}{\s} \; (1+\ms^2) \;| C_7^{\mbox{eff}} |^2 
		- 4  {\rm Re} \left[ ( \cnt )^\ast  C_7^{\mbox{eff}} \right]
		\right\} , \\
        {{T_3}^{(g)}}^{L/R} & = & \frac{-1}{x^2} \mb \, \lto
                \left\{ | \cnt |^2 + \frac{4}{\s^2} 
		\left[ 2 \, (\z) - \s \right] \; (1-\ms^2)\; | C_7^{\mbox{eff}} |^2 
\right.
\nonumber \\
& & \left.
		+ \frac{4}{\s} \; (1+ \ms^2) \; {\rm Re} \left[ ( \cnt )^\ast \, C_7^{\mbox{eff}} \right]
		\right\} \, , \\
	{{T_1}^{(\delta)}}^{L/R} & = & 
		\frac{1}{2} \, \mb \, ( \loo + 3 \lto ) \left\{
		\left[ \frac{1}{x} - \frac{2}{x^2} 
			\left( 1 - \z \right)^2 \right] 
			| \cnt |^2
		\right.
		\nonumber \\
	& &  
		- \frac{4}{\s^2 x^2} \left[
-2 \s -6 \ms^2 \s + \s x + \ms^2 \s x +  4 \ms^2 \s \z + 4 \z^2 + 4 \ms^2 \z +2 \s \z^2
\right.
\nonumber \\
& & \left. + 2 \ms^2 \s \z^2-2 x \z^2 -2 \ms^2 x \z^2 -4 \z^3 -4 \ms^2 \z^3
		\right] | C_7^{\mbox{eff}} |^2 
		\nonumber \\
	& & 
		- \frac{4}{\s x^2} \left[ 
-2 \s + 2 \z -2 \ms^2 \z + 2 \s \z -x \z -2 \z^2 
\right.
\nonumber\\
& & \left. \left.
+ 2 \ms^2 \z^2
		\right] {\rm Re} \left[ ( \cnt )^\ast \, C_7^{\mbox{eff}} \right]
		\right\} \, , \\
        {{T_2}^{(\delta)}}^{L/R} & = &
		\mb \, ( \loo + 3 \lto ) \left[
		\frac{1}{x} - \frac{2}{x^2} \left( 1 - \z \right)
		\right] \left[ 
		| \cnt |^2 - \frac{4}{\s}\; (1+\ms^2)\; | C_7^{\mbox{eff}} |^2 \right] \, , \\
        {{T_3}^{(\delta)}}^{L/R} & = & 
		\mb \, ( \loo + 3 \lto ) \left\{
		- \frac{1}{x^2} \left( 1 - \z \right) | \cnt |^2 
		\right.
		\nonumber \\
	& & \left. 
		\; \; \; \; \; \; \; \; \; \; 
		+ \frac{4}{\s^2} \left[
		\frac{1}{x} \z - 
		\frac{1}{x^2} \left( 1 - \z \right)
			\left( 2 \, (\z) - \s \right) \right]\; (1-\ms^2)\; | C_7^{\mbox{eff}} |^2
		\right.
		\nonumber \\
	& & \left. 
		\; \; \; \; \; \; \; \; \; \; 
		- \frac{2}{\s x^2} \left[
2+2 \ms^2 -x -2 \z -2 \ms^2 \z
		\right] {\rm Re} \left[ ( \cnt )^\ast \, C_7^{\mbox{eff}} \right]
		\right\} \, . 
\end{eqnarray}
In the above expressions, the variable $x$ is defined as
 $x \equiv 1 + \s - 2 \, (\z) - \ms^2 + i \, \epsilon $.

%
%
%
%
\section{Auxiliary functions $E_1(\hat{s}, \hat{u})$ and
 $E_2(\hat{s}, \hat{u})$ in the Dalitz distribution $d^2{\cal 
B}/d\s d\u (b \to s \ell^+ \ell^-)$  in the HQE Approach}

 \setcounter{equation}{0}

In this appendix we give 
the auxiliary functions $E_1 (\s, \u)$
and $E_2 (\s, \u)$, multiplying the
delta-function $\delta [\u (\s,\hat{m}_s) -\u^2]$
and its first derivative 
$\delta^\prime [ \u (\s,\hat{m}_s) -\u^2 ]$, respectively,
appearing in the power corrected Dalitz distribution in \bsll
given in eq.~(\ref{eqn:dddw}) in the text.
 
\begin{eqnarray}
	E_1 (\s, \u) & = & 
		\frac{1}{3} \left\{ 
		2 \, \lo \left[ 1 -4 \ms^2+6 \ms^4-4 \ms^6 + \ms^8 -2 \ms^2 \s + 4 \ms^4 \s -2 \ms^6 \s + 2 \ms^2 \s^3  - \s^4
\right. \right.
\nonumber \\
& & \left. \left.
+ \u^2 \left( 1 -2 \ms^2 + \ms^4 -2 \ms^2 \s + 4 \, \s + \s^2 \right) \right]
	\right.
	\nonumber \\
& &	 \left. 
		+ 3 \, \lt \; (1-\ms^2+\s) \left[ -1 + 7 \ms^2 -11 \ms^4 +5 \ms^6 + 11 \, \s +10 \ms^2 \s -5 \ms^4 \s - 15 \, \s^2 \right. \right.
\nonumber \\
& &\left. \left.
-5 \ms^2 \s^2 +5 \s^3 
		+ \u^2 \left( 1 -5 \ms^2 +5 \, \s  \right) \right]
		\right\}
	\nonumber \\
	& & 
	\times	\left( |C_9^{\mbox{eff}}|^2 + |C_{10}|^2 \right)
	\nonumber \\
	& &	+ \frac{4}{3 \, \s} \left\{
		2 \, \lo \left[ 1 -3 \ms^2+2 \ms^4+2 \ms^6 -3 \ms^8+\ms^{10} -10 \ms^2 \s + 18 \ms^4 \s -6 \ms^6 \s -2 \ms^8 \s
\right. \right.
\nonumber \\
& &
 +16 \ms^4 \s^2 -6 \ms^2 \s^3 +2 \ms^4 \s^3 - \s^4 -\ms^2 \s^4 
\nonumber \\
& & \left. 
- \u^2 \left( 1 - \ms^2 - \ms^4 +\ms^6  + 4 \s + 2 \ms^2 \s - 2 \ms^4 \s + \s^2+ \ms^2 \s^2 \right)  \right]
	\nonumber \\
& &	 
		+ 3 \, \lt  \; (1-\ms^2+\s) \left[ 3 +2\ms^2-8 \ms^4 -2 \ms^6 +5 \ms^8+ 3 \s - 35 \ms^2 \s -27 \ms^4 \s -5 \ms^6 \s
\right.
\nonumber \\
& &
  -11 \s^2+8 \ms^2 \s^2 -5 \ms^4 \s^2  +5 \s^3 +5 \ms^2 \s^3
\nonumber \\
& & \left. \left.
 + \u^2 \left( 
			3 +8 \ms^2+5 \ms^4 - 5 \s-5 \ms^2 \s  \right) \right]
		\right\} |C_7^{\mbox{eff}}|^2
	\nonumber \\
	& &	+ 8 \left\{ \frac{2}{3} \lo (1 -4 \ms^2 +6 \ms^4 - 4 \ms^6+\ms^8 -\s-\ms^2 \s +5\ms^4 \s -3 \ms^6 \s  + \s^2 + 3\ms^4 \s^2
\right.
\nonumber \\
& &
 -\s^3- \ms^2 \s^3) 
\nonumber \\
 & & \left.
		+ \lt (1 -\ms^2 + \s) \left[ 4-3 \ms^2 -6 \ms^4+5 \ms^6 -6 \s-4 \ms^2 \s -10  \ms^4 \s +2 \s^2+5 \ms^2 \s^2 + \u^2	\right]
		\right\} 
\nonumber \\
& &
	\, Re(C_9^{\mbox{eff}}) \, C_7^{\mbox{eff}}
	\nonumber \\
	& &	+ 4 \, \s \, \u \left[ - \frac{4}{3} \lo \, \s 
		+ \lt \left( 7 -2 \ms^2 -5 \ms^4 + 2 \, \s +10 \ms^2 \s - 5 \, \s^2 \right) \right] 
		\, Re(C_9^{\mbox{eff}}) \, C_{10}
	\nonumber \\
	& &	+ \frac{8}{3} \u \left[ -4 \, \lo \, \s \; (1+\ms^2)
\right.
\nonumber \\
& &  \left. + 
	3  \lt \left( 5 + \ms^2 - \ms^4 -5 \ms^6+ 2  \s +4 \ms^2 \s+10 \ms^4 \s  - 3 \, \s^2-5 \ms^2 \s^2 \right) \right]
		 {C_{10}}^\ast \, C_7^{\mbox{eff}},
		\label{eqn:e1} 
\end{eqnarray}
\begin{eqnarray}
	E_2 (\s, \u) & = & 
		\frac{2}{3} \lo \left( 1 - \ms^2 + \s \right)^2 \u(\s,\ms)^2
\left[  
			\left( 1 -2 \ms^2 +\ms^4 - \s^2 - \u^2 \right)
                   	\left( |C_9^{\mbox{eff}}|^2 + |C_{10}|^2 \right)
\right.
	\nonumber \\
	& & 
  + 4	\left( 1 - \ms^2 -\ms^4+\ms^6-8 \ms^2 \s - \s^2-\ms^2 \s^2 + \u^2+\ms^2 \u^2 \right)
        \frac{|C_7^{\mbox{eff}}|^2}{\s}
	\nonumber \\
	& &  
		+ 8  \left( 1 -2 \ms^2+\ms^4 - \s -\ms^2 \s \right) 
			Re(C_9^{\mbox{eff}}) \,C_7^{\mbox{eff}}
\nonumber \\
 & & \left.
	+ 4 \, \s \, \u \; 
			Re(C_9^{\mbox{eff}}) \,C_{10}
		+ 8 \, \u \; (1+\ms^2) \;
						Re(C_{10}) \, C_7^{\mbox{eff}} 
\right]		
\nonumber \\
\, .
		\label{eqn:e2} 
\end{eqnarray}
%
%
%
\section{The decay rate $\Gamma(b \to s \ell^+ \ell^-)$ in the 
$V-A$ limit and comparison with the existing results}
 \setcounter{equation}{0}
In this appendix we compare our results for the power corrections
in the decay \bxsll  
with the  ones for the decays $B \to X_c \ell \nu_\ell$,  
derived by Manohar and Wise (MW) \cite{manoharwise}.
In doing this, we shall reduce the matrix element for the
decay \bxsll to the one encountered in
$B \to X_c \ell \nu_\ell$, obtained by the replacements:
\begin{eqnarray}
        C_9 & = & - C_{10} = \frac{1}{2}
                \, , \\
        C_7 & = & 0 
                \, , \\
        \left( \frac{G_F \, \alpha}{\sqrt{2} \, \pi}
                V_{ts}^\ast V_{tb}\right)
                & \rightarrow &
        \left( - \frac{4 \, G_F}{\sqrt{2}} V_{cb} \right)
                \, .   
\end{eqnarray}
This amounts to keeping only the CC $V-A$ contribution in \bxslll. 

We remark that our hadronic states and the ones used by Manohar and 
Wise are differently normalized, with the two related by
\begin{equation}
	\left| M \right> = \sqrt{2 \, M_B} \, 
		\left| M \right>^{MW} 
		\, .
\end{equation}

Hence, the forward scattering amplitudes are related through,
\begin{equation}
        {T_{\mu \nu}}  =  - {1 \over 2M_B} {T_{\mu \nu}}^{MW}
                \, . 
\end{equation}

Likewise, the matrix elements of the kinetic energy and
the magnetic moment operators in the HQE 
approach are related,
\begin{eqnarray}
	\loo & = &  - 2 \, {m_b}^2 \, K_b 
		\, , \\
	3 \, \lto & = &  - 2 \, {m_b}^2 \, G_b 
		\, , \\
	\loo + 3 \, \lto & = & - 2 \, {m_b}^2 \, E_b 
		= - 2 \, {m_b}^2 \, ( K_b + G_b ) 
		\, . 
\end{eqnarray}
Note further that  our structure functions $T_{i}$ are dimensionless, as 
opposed to the ones employed in \cite{manoharwise}. Thus,
 \begin{eqnarray}
	\left( T_1 \right)^{MW} & = & 
		-{ {  \left( T_1 \right) } \over {2M_B}} 
		\, , \\
	\left( T_2 \right)^{MW} & = & 
		-{ { \left( T_2 \right) } \over {2M_B}}
     		\, , \\
	\left( T_3  \right)^{MW} & = & 
		-{ {\left( T_3 \right) } \over {2M_B m_b}}
		\, .
\end{eqnarray}
With these replacements, the structure functions $T_{i}, ~i=1,2,3$
given in the text and Appendix A in this paper agree with those in MW up to 
the indicated normalization factors in the $V-A$ limit.
Note that the function  
$\Delta_0$ defined in eq.~(3.4) of \cite{manoharwise} transcribes to 
$\Delta_0 = {m_b}^2 \, x$ in our notation, with $x$ defined in Appendix A.

 After integrating  $T_{\mu \nu} \, L^{\mu \nu}$ in the 
complex $\z$ plane, we have also compared the double differential
distribution $(1/\Gamma_b) {\rm d} \Gamma/{\rm d}y {\rm d}{\hat{q}^2}$,
given in eq.~(5.2) of \cite{manoharwise}. Taking into account that our
differential distributions are defined in terms of the variables 
$\u$ and $\s$, as opposed to the variables  $\s$ and $y$ with 
$y \equiv 2 \, \hat{E}_e$ used in \cite{manoharwise}, and
making the variable transformation $y = \u - 2 \, \z $, 
we reproduce their result.

Finally, using the correspondence (C-1) - (C-3), the differential 
dilepton distribution in \bsll reduces to (with $m_s =0$)
\begin{equation}
\frac{{\rm d}\Gamma}{{\rm d}\s} =  \Gamma^{b}
               \left( \frac{1}{3} (1-\s)^2 (1+2 \s) \; (2 + \lo)
                + ( 1 - 15  \s^2 + 10 \s^3) \lt \right) ~,
\end{equation}             
which, on integration gives
\begin{equation}
	 \Gamma = 
		\Gamma^b \, \left( 1 + \frac{1}{2} \lo
		- \frac{9}{2} \, \lt \right) 
		\, , 
		\label{eqn:gmw}
\end{equation} 
where $\Gamma^b$ is the parton model decay width. The above expression
agrees with the well known result of Bigi et al \cite{georgi}. 
Doing the
same manipulation on the corresponding expressions by FLS
\cite{falketal}, we get instead
\begin{equation}	
	\Gamma^{FLS}  =  \Gamma^b \,
 \left( 1 + \frac{17}{3} \lo + 13 \, \lt \right) 
		\, ,
\end{equation} 
where in $\Gamma^{FLS}$ also 
only the $V-A$ contributions are kept. This disagrees with our result as 
well as with the one in \cite{georgi}. 
%
%
%
\section{Equivalence of FB Asymmetry and Energy Asymmetry in \bxslll}

\setcounter{equation}{0}
        
In this appendix we address a peripheral issue, namely  that the 
quantity Energy asymmetry, introduced 
in \cite{choetal}, is simply related to the FB asymmetry, defined in
\cite{amm91,agm94}, and is not an independent quantity.
It is easy  to show that the configuration in which
$l^+$ is scattered in the forward direction, measured with respect to the
direction
of the $B$-meson momentum in the dilepton c.m.s.,  corresponds
to the events in which $E_{-} > E_{+)}$ in the $B$-meson rest
frame, where $E_{\pm}$ represents the $\ell^\pm$-energy. To that end let us 
suppose 
that in the dilepton c.m.s., $l^+$ is scattered in the forward direction.
In this frame, the four-momenta of $\ell^{+}$ and $\ell^{-}$ are given as
\begin{eqnarray}
        p_\mu^+ & = & \left( \epsilon, p_\parallel , p_\perp \right)
                \, , \\
        p_\mu^- & = & \left( \epsilon, -p_\parallel , -p_\perp \right)  
                \, ,
\end{eqnarray}
where $p_\parallel > 0$ is the longitudinal momentum and $p_\perp$ is
the transverse momentum. Boosting the momenta with the velocity of $B$
meson ($v$) takes one to the $B$-meson rest frame, where $E_{+}$ and $E_{-}$,
 are given by
\begin{eqnarray}
        E_+ & = & \epsilon \, \frac{1}{\sqrt{1 - v^2}}
                - p_\parallel \, \frac{v}{\sqrt{1 - v^2}}
                \, , \\
        E_- & = & \epsilon \, \frac{1}{\sqrt{1 - v^2}}
                + p_\parallel \, \frac{v}{\sqrt{1 - v^2}}
                \, .
\end{eqnarray}
implying that for forward scattered $\ell_{+}$ in the dilepton c.m.s.,
one has $E_{+} < E_{-}$ in the $B$ meson rest frame.
 By using the definition of the
FB-asymmetry in \cite{amm91,agm94}, we obtain the following simple
relation to the energy asymmetry of \cite{choetal},
\begin{equation}
        \frac{{\rm d}{\cal A} (\s)}{{\rm d}\s} = \int_0^1
                \frac{{\rm d}^2 {\cal B}}{{\rm d}\s \, {\rm d}z} \, {\rm d}z
                - \int_{-1}^{0}
                \frac{{\rm d}^2 {\cal B}}{{\rm d}\s \, {\rm d}z} \, {\rm d}z
        \, ,
\end{equation}
where $z \equiv \cos \theta$,
\begin{equation}
        \int \, \frac{{\rm d}{\cal A}(\s)}{{\rm d}\s} \, {\rm d}\s =
                {\cal B} \times A
                \, ,
\end{equation}
where $A \equiv {\left[ N (E_- > E_+) - N (E_+ > E_-) \right]}/{
\left[ N (E_- > E_+) + N (E_+ > E_-) \right]}$
is the energy asymmetry defined in \cite{choetal}. Hence $A$ of 
\cite{choetal} is identical to the normalized FB asymmetry $\overline{A}$
calculated in this paper.
That the two quantities are related can also be seen by writing the
 Mandelstam variable $\u$, defined previously,
in the dilepton c.m.s. and the $B$-meson rest frame:
\begin{eqnarray}
        \u & = & - \u (\s) \, \cos \theta
                \nonumber \\
           & = & 2 ( \hat{E}_+ - \hat{E}_- ) \, .
\end{eqnarray}
%
%
%
\section{ Dalitz distribution $d^2\Gamma(B \to X_s \ell^+ 
\ell^-)/ds du$ and FB asymmetry in the Fermi motion model}

 \setcounter{equation}{0}
\begin{figure}[htb]
\vskip -2.0truein
\centerline{\epsfysize=7in
{\epsffile{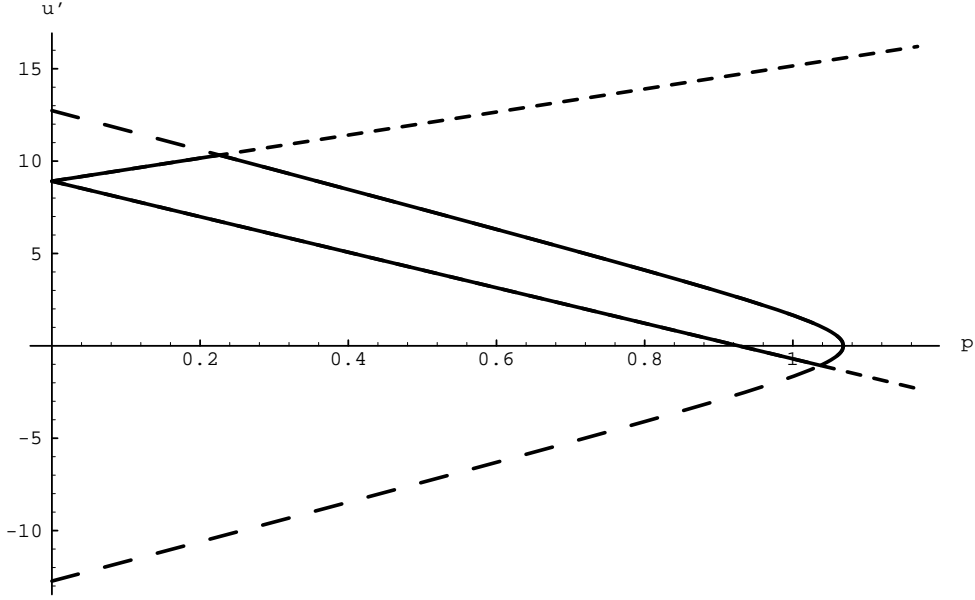}}}
\vskip -2.0truein
\caption[]{Phase space boundaries for the $u^\prime$ and $p$ integrations 
with fixed values of $s$ and $u$ drawn for $s=15$ GeV$^2$ and
$u= 8.9$ GeV$^2$. The integration region (solid 
curve) is given by the intersection of $u^\prime_\pm$ (short dashed)
and $\pm u(s,p)$ (long dashed curve). The fermi motion parameters used are
$(p_F,m_q)=(450,0)$.}
\label{fig:pspace}
\end{figure}

We start with the differential decay rate
 $d^3\Gamma_B/{\rm d}s \, {\rm d}u \, {\rm d}p$, describing the decay
\bsll of a moving $b$-quark having a mass $W(p)$ and 
three momentum $\vert p \vert \equiv p$ 
with a distribution $\phi(p)$, which will be
taken as a Gaussian \cite{aliqcd},
\begin{equation}
	\frac{{\rm d}\Gamma_B}{{\rm d}s \, {\rm d}u \, {\rm d}p} = 
	\int^{u^\prime_{\rm max}}_{u^\prime_{\rm min}} \, {\rm d}u^\prime \, 
	\frac{{W(p)}^2}{M_B} \, p \, \phi(p) \, 
	\frac{1}{\sqrt{{u^\prime}^2 + 4 \, {W(p)}^2 \, s}} \, 
	\left[ \frac{{\rm d}^2 \Gamma_b}{{\rm d}s \, {\rm d}u^\prime} 
	\right] \; . 
\end{equation}
Here, ${\rm d}^2 \Gamma_b/{\rm d}s \, {\rm d}u^\prime$
is the  double differential decay rate of a b-quark at rest and
can be written in the case of \bsll as,  
\begin{equation}
	\frac{{\rm d}^2 \Gamma_b}{{\rm d}s \, {\rm d}u^\prime} =  
	\left | V_{ts} \, V_{tb} \right|^2 \, 
	\frac{{G_F}^2}{192 \, \pi^3} \, \frac{1}{{W(p)}^3} \, 
	\frac{3 \, \alpha^2}{16 \, \pi^2} \left[ F_1(s,p) +  
	F_2(s,p) \, u^\prime + F_3(s,p) \, {u^\prime}^2 
	\right] \; , 
\end{equation}
and the three functions have the following expressions,
\begin{eqnarray}
	F_1(s,p) & = & \left[ \left( {W(p)}^2 - {m_s}^2 \right)^2 - s^2 \right]
		\left( |{C_9^{\mbox{eff}}}|^2 + |{C_{10}}|^2 \right) 
		\nonumber \\
	& &  
		+ 4 \left[ {W(p)}^4 - {m_s}^2 \, {W(p)}^2 - {m_s}^4 
			+ \frac{{m_s}^6}{{W(p)}^2} - 8 \, s \, {m_s}^2 
			- s^2 \, \left( 1 + \frac{{m_s}^2}{{W(p)}^2} \right) 
			\right]	\, \frac{{W(p)}^2}{s} \, |{C_7^{\mbox{eff}}}|^2 
		\nonumber \\
	& &  
		- 8 \left[ s \, \left( {W(p)}^2 + {m_s}^2 \right) 
			- \left( {W(p)}^2 - {m_s}^2  \right)^2 \right] \, 
			{\rm Re} (C_7^{\mbox{eff}} \, {C_9}^{\mbox{eff}})
			\; , \\
	F_2(s,p) & = & 4 \, s \, {\rm Re}({C_9}^{\mbox{eff}} \, C_{10}) + 
		8 \, \left( {W(p)}^2 + {m_s}^2 \right) \, C_{10} \, C_7^{\mbox{eff}} 
		\; , \\
	F_3(s,p) & = & - \left(|{{C_9}^{\mbox{eff}}}|^2 + |{C_{10}}|^2 \right) 
		+ 4 \left[ 1 + \left( \frac{{m_s}^2}{{W(p)}^2} \right)^2 
		\right] \, \frac{{W(p)}^2}{s} \, |{C_7^{\mbox{eff}}}|^2  \; , 
\end{eqnarray}
which can be read off directly from eq.~(\ref{eqn:dddw})
 in the limit $\lambda_{i} =0;~i=1,2$.
Note that the Wilson coefficient ${C_9}^{\mbox{eff}}$ also has an
implicit $W(p)$ dependence, as can be seen in the text.
The integration limit for $u^\prime$ is determined through the equations 
\begin{eqnarray}
	u_{\rm max}^\prime & \equiv & {\rm Min} \, 
		\left[ u_+^\prime, u(s,p) \right] \; , \\
	u_{\rm min}^\prime & \equiv & {\rm Max} \, 
		\left[ u_-^\prime, -u(s,p) \right]	\; ,
\end{eqnarray}
where 
\begin{equation}
	u^\prime_{\pm} \equiv \frac{E_W}{M_B} \, u \pm 
	\frac{p}{M_B} \, \sqrt{4 \, s \, {M_B}^2 + u^2} \; , 
\end{equation}
\begin{equation}
E_W=\sqrt{W(p)^2+p^2} \; ,
\end{equation}
and
\begin{equation}
	u(s,p) \equiv \sqrt{\left[ s- \left( W(p) + {m_s} \right)^2  \right] 
		\left[ s -\left( W(p) - {m_s} \right)^2 \right]} \; . 
\end{equation}

A typical situation in the phase space is displayed in Fig.~\ref{fig:pspace}.
 Integration over $p$ gives the double differential 
decay rate (Dalitz distribution) including the Fermi motion. The result is, 
\begin{eqnarray}
	\frac{{\rm d}^2 \Gamma_B}{{\rm d}s \, {\rm d}u} & = & 
	  \left | V_{ts} \, V_{tb} \right|^2 \, 
	\frac{{G_F}^2}{192 \, \pi^3} \, 
	\frac{3 \, \alpha^2}{16 \, \pi^2} 
	\int_0^{p_{\rm max}} {\rm d}p  \, 
	\frac{1}{W(p) 2 M_B} \, p \, \phi(p) \,  
	 \nonumber \\ 
	& & \; \; \; \; \; \; 
	\left\{ F_1(s,p) \, \log \left| \frac{u_{\rm max}^\prime + 
		\sqrt{{u_{\rm max}^\prime}^2 + 4 \, {W(p)}^2 \, s}}{
		u_{\rm min}^\prime + 
		\sqrt{{u_{\rm min}^\prime}^2 + 4 \, {W(p)}^2 \, s}} \right| 
	\right. \nonumber \\ 
	& & \; \; \; \; \; \; \left.
	+ F_2(s,p) \, \left[ \sqrt{{u_{\rm max}^\prime}^2 + 4 \, {W(p)}^2 \, s} 
		- \sqrt{{u_{\rm min}^\prime}^2 + 4 \, {W(p)}^2 \, s} \right]
	\right. \nonumber \\ 
	& & \; \; \; \; \; \; \left.
	+ F_3(s,p) \, \frac{1}{2} \, \left[ 
	u_{\rm max}^\prime \, \sqrt{{u_{\rm max}^\prime}^2 + 4 \, {W(p)}^2 \, s} 
	- u_{\rm min}^\prime \, \sqrt{{u_{\rm min}^\prime}^2 + 4 \, {W(p)}^2 \, s} 
	\right. \right. \nonumber \\ 
	& & \; \; \; \; \; \; \; \; \; \; \; \; \; \; \; \; \; \; \left. \left.
	- 4 \, {W(p)}^2 \, s \, 
	\log \left| \frac{u_{\rm max}^\prime + 
		\sqrt{{u_{\rm max}^\prime}^2 + 4 \, {W(p)}^2 \, s}}{
		u_{\rm min}^\prime + 
		\sqrt{{u_{\rm min}^\prime}^2 + 4 \, {W(p)}^2 \, s}} \right| 
	\right] \right\} \; .
\end{eqnarray}
Note that the upper limit in $p$ integration, $p_{\rm max}$ is 
determined such that $p$ satisfies, 
\begin{equation}
	u_{\rm max}^\prime(p_{\rm max},s,u) = 
		u_{\rm min}^\prime(p_{\rm max},s,u) \; .
\end{equation}
Lastly, the normalized differential FB asymmetry including the Fermi motion becomes, 
\begin{equation}
	\frac{{\rm d}\overline{\cal A}}{{\rm d}s} = 
	\frac{\int_{-u_{\rm ph}}^0 \frac{{\rm d}\Gamma_B}{
		{\rm d}s \, {\rm d}u} \, {\rm d}u 
	- \int^{u_{\rm ph}}_0 \frac{{\rm d}\Gamma_B}{
		{\rm d}s \, {\rm d}u} \, {\rm d}u 
	}{\int_{-u_{\rm ph}}^0 \frac{{\rm d}\Gamma_B}{
		{\rm d}s \, {\rm d}u} \, {\rm d}u 
	+ \int^{u_{\rm ph}}_0 \frac{{\rm d}\Gamma_B}{
		{\rm d}s \, {\rm d}u} \, {\rm d}u } \; , 
\end{equation}
where 
\begin{equation}
	u_{\rm ph} \equiv \sqrt{
		\left[ s - \left( M_B + M_X \right)^2 \right]
		\left[ s - \left( M_B - M_X \right)^2 \right]} \; ,
\end{equation}
and 
\begin{equation}
	M_X \equiv {\rm Max} \left[ m_K, m_s + m_{\rm q} \right] \; , 
\end{equation}
with $m_{\rm q}$ the spectator quark mass and $m_K$ the 
kaon mass. Since the calculations are being done for an inclusive decay
\bxslll, we should have put this threshold higher, say starting from 
$m_K + m_\pi$, but as this effects the very end of a steeply falling 
dilepton mass spectrum, we have kept the threshold in \bxsll at 
$m(X_s) =m_K$.
\end{appendix}

\end{document}